\documentclass{article}

\usepackage[english]{babel}
\usepackage{amsmath}
\usepackage{amssymb}
\usepackage{amsthm}
\usepackage{a4wide}
\usepackage[dvips]{graphicx}
\usepackage{url}
\usepackage{psfrag}
\usepackage{subfig}

\title{The camera method, or how to track numerically a deformable particle moving in a fluid network.}

\author{Baptiste Moreau$\phantom{}^1$, Philippe Dantan$\phantom{}^1$, Patrice Flaud$\phantom{}^1$ and Benjamin Mauroy$\phantom{}^{1,2,\star}$\\
\\
{\small $\phantom{}^1$ Laboratoire Mati\`ere et Syst\`emes Complexes,}\\
{\small UMR CNRS 7057, Universit\'e Paris 7 Denis Diderot, Paris, France.}\\\\
{\small $\phantom{}^2$ Laboratoire J.A. Dieudonn\'e,}\\
{\small UMR CNRS 7351, Universit\'e de Nice-Sophia Antipolis, Nice, France.}\\\\
{\small $\phantom{}^{\star}$ Corresponding author, email: benjamin.mauroy@unice.fr}}

\date{}

\begin{document}

\maketitle

\noindent {\bf Corresponding author :}\\
Benjamin Mauroy\\
Laboratoire J.A. Dieudonn\'e, Universit\'e de Nice-Sophia Antipolis,\\
Parc Valrose, 06108 Nice cedex 2,\\
France.\\
email: benjamin.mauroy@unice.fr\\
tel: +33 (0) 4 92 07 62 10\\
fax: +33 (0) 4 93 51 79 74\\

\newpage

\noindent{\bf Abstract}\\\\
The goal of this work is to follow the displacement and possible deformation of a free particle in a fluid flow in 2D axi-symmetry, 2D or 3D using the classical finite elements method without the usual drawbacks finite elements bring for fluid-structure interaction, i.e. huge numerical problems and strong mesh distortions. Working with finite elements is a choice motivated by the fact that finite elements are well known by a large majority of researchers and are easy to manipulate. The method we describe in this paper, called the camera method, is well adapted to the study of a single particle in a network and most particularly when the study focuses on the particle behaviour. The camera method is based on two principles: 1/ the fluid structure interaction problem is restricted to a neighbourhood of the particle, thus reducing drastically the number of degrees of freedom of the problem; 2/ the neighbourhood mesh moves and rotates with the particle, thus avoiding most of the mesh distortions that occur in a standard ALE method. In this article, we present the camera method and the conditions under which it can be used. Then we apply it to several examples from the literature in 2D axi-symmetry, 2D and 3D.\\




\section{Introduction}

Tracking a particle in a fluid has many applications in a wide range of up to date thematics: biofluidics and medicine (red blood cells, drugs delivery, etc.), electrophoresis, magnetic particle driving, aerosols, pollutants, etc. The particle can either be a solid, a deformable particle or a fluid enclosed inside a membrane (vesicle or capsule). It can move in a fluid domain whose size is either of the scale of the particle (a red blood cell in a capillary) or many scales larger than the particle (a red blood cell in an artery, an aerosol in the lungs, a pollutant in a house, etc.).

\begin{figure}[h!]
\centering
\bf{A}\includegraphics[height=4cm]{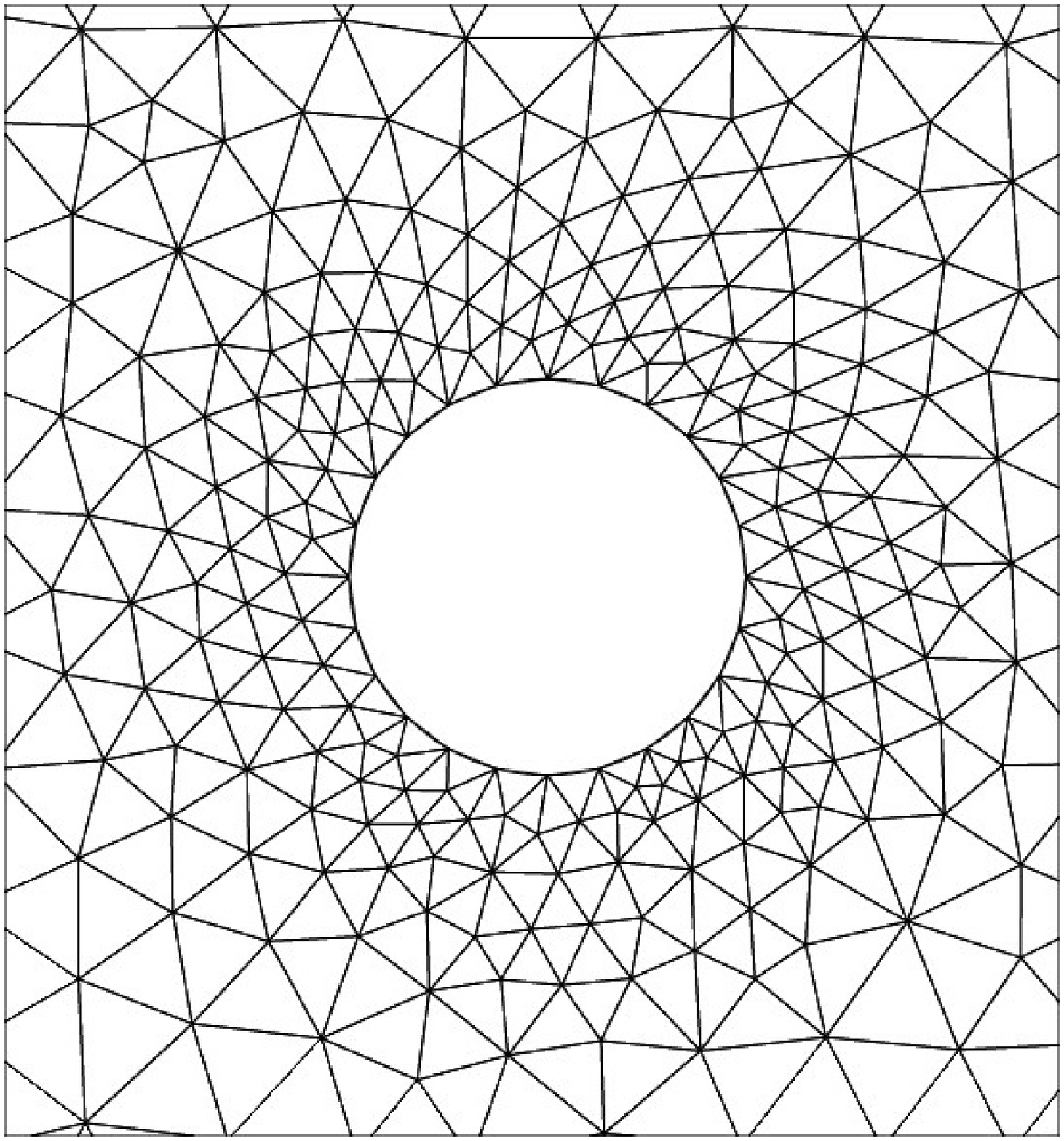}
\bf{B}\includegraphics[height=4cm]{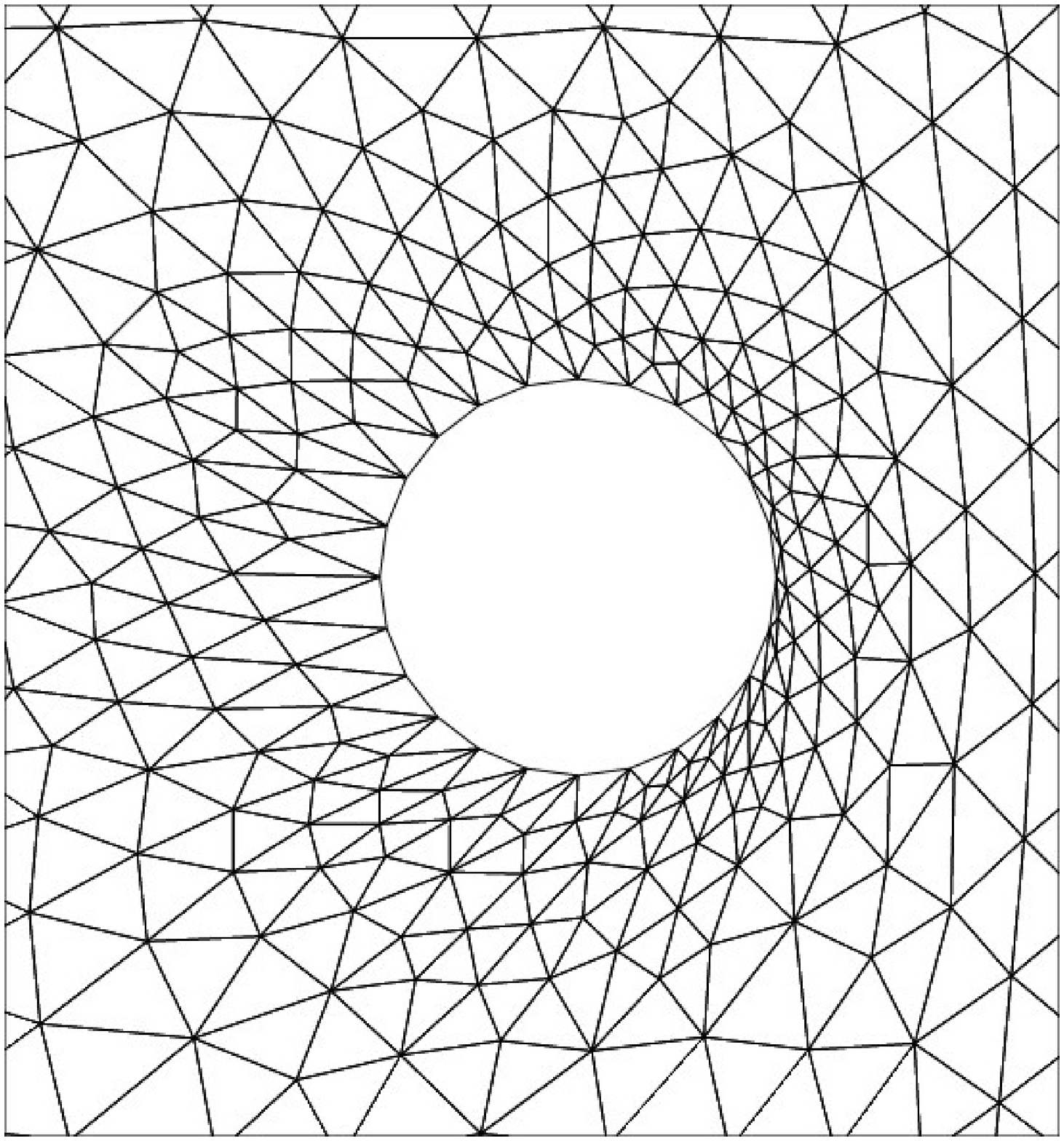}
\bf{C}\includegraphics[height=4cm]{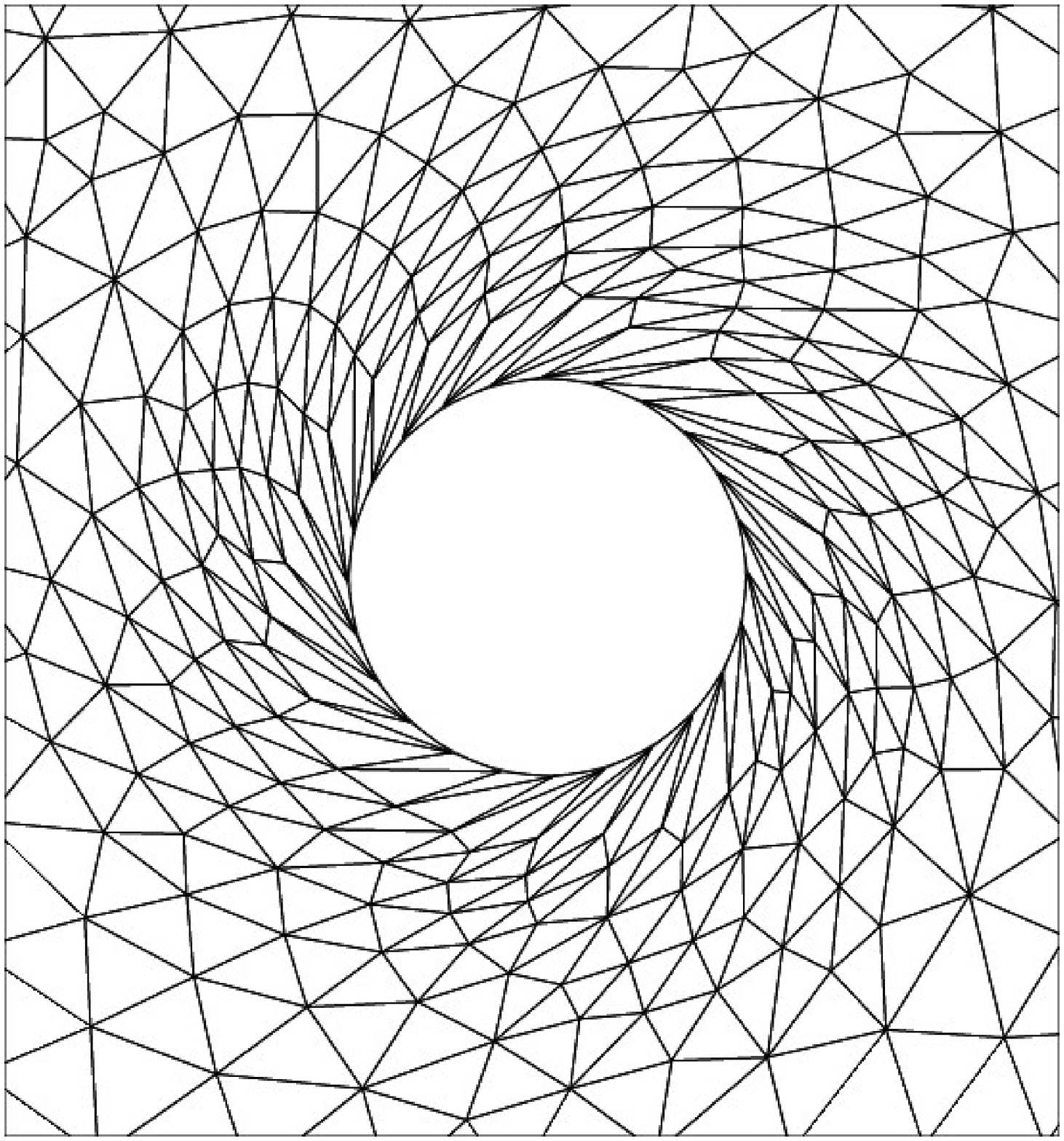}
\caption{Examples of the distortion of a mesh (A) due to translation (B) and rotation (C).}
\label{EFdist}
\end{figure}

A typical application is the study of an isolated vesicle in external flows. This subject is extremely challenging, since vesicles exhibit complex behaviours depending on a small number of physical parameters (reduced volume, internal/external viscosity ratio, capillary number). 
Behaviours of vesicles such as tank-threading, tumbling or vacillating-breathing appears when scanning the range of these parameters and all have been predicted by theory \cite{Misbah,Misbah2,Skalak} and observed in experiments \cite{Gold,Deschamps}. To improve the understanding of these phenomena, numerical simulations are very useful. Different computational methods are used: the boundary integral method \cite{Kaoui, Kraus, Veera1,Veera2, Veera3}, particles-based methods \cite{Richardson} and hybrid methods \cite{Noguchi,Walter}. Each method models the fluid and/or vesicle behaviour in a particular way and accuracy often goes with high computational costs. Very few studies use the classical finite elements method, except for research of stationary shapes \cite{Feng, Ma} or for studies limited in 2D axi-symmetry for non stationary regime \cite{MauroyRBC}. The major reason is that, although this method is well known and spread in laboratories, it is not, at first sight, well adapted to such problems, be it vesicles or more generally a particle in an external flow. Actually, if one wants to use standard finite elements, an ALE (Arbitrary Lagrangian Eulerian) method will be generally used. The particle displacements are solutions of Lagrangian equations in the reference frame (mechanics equations) while fluid velocities and pressures are solutions of Eulerian equations in the deformed frame (incompressible Navier-Stokes or Stokes equations). The deformed frame is determined thanks to a transformation of the reference frame. This coordinates change coincides with the particle displacement on the particle subdomain and is the results of a bearing (i.e. is extended) on the fluid subdomain. Such as, this method needs the whole fluid region and particle to be meshed. The first difficulty appears when particle and fluid domain scales are different because the mesh elements will be inhomogeneous and in large numbers in order to cover the range of scales. If this problem can anyway be handled, ALE method will work well as long as the solid displacements remain small comparatively to the domains size. Indeed, for large displacements (even in the case where particle deformations are small), the definition of the transformation from the reference frame to the deformed frame becomes less good (the determinant of its Jacobian shifts away from $1$). This situation brings convergence problems in numerical simulations: it results in mesh distortions, such as in the examples shown on figure \ref{EFdist}. Eventually the convergence is not any more possible although the physical problem remains valid. A simple translation or rotation of the particle could cause mesh distortions. A solution commonly used to solve this situation is to remesh the system regularly. Nevertheless, the computation of a new mesh can be time consuming and the projection of the previous solution on the new mesh generates additional errors. This is the reason why most studies using finite elements are limited to the research of stationary shapes.

In this work we propose a new method based on the classical ALE method. The fluid domain that is meshed and simulated in our approach is reduced to a close neighbourhood of the particle. Moreover, we make the mesh of the neighbourhood follow the particle and rotate with it. The number of mesh elements is thus limited because only a small part of the fluid domain is meshed and mesh deformation due to translations and rotations are suppressed. As such, the subdomain on which we propose to work can be considered as the frame of a camera that is attached to the particle in the fluid flow, therefore we called the method "camera method". The drawback is that we must determine approximations of fluid conditions on the boundary of the camera, either by analytical means (for example if the flow is in Poiseuille's regime) or by a preliminary simulation of the fluid in the domain without the particle.

The underlying hypothesis in the work presented in this paper is that the particle does not influence the fluid velocities outside the camera frame. In section \ref{discussion}, we discuss the consequences and limitations of this hypothesis and we give hints on how the camera method can be extended when this hypothesis is not verified.

We will first describe the numerical method in detail with the different equations involved and the specificity of each dimension (2D axi-symmetry, 2D and 3D). Then we will apply our method to several examples of various dimensions.

\section{Numerical method of camera}

\subsection{Principles}

We consider a particle in a fluid network. Our goal is to follow the particle along the network with numerical simulations and we want to simulate only the deformation of the particle and the fluid velocities and pressures in a neighbourhood of the particle, i.e. in the camera frame.

The camera frame is typically a sphere centred on the gravity centre of the particle. The camera frame contains the particle surrounded by fluid. The fluid-structure interaction problem in the camera frame is similar to a classical fluid-structure interaction problem, except for the transformation from the reference frame to the deformed frame. Actually, the transformation in the fluid is the solution of a partial derivative equation (to define the bearing) and in order to solve the equation, the displacements on the boundary must be given. The displacements on the boundary of the particle are solutions of the mechanics equations; the displacements on the boundary of the camera frame depend also on the solution of the mechanics equations since they follow the global displacement of the particle (translation and rotation(s)).

To have correct fluid velocities and pressures inside the camera, we must also define correct fluid conditions on the boundary of the camera. We choose to impose Dirichlet velocity conditions on the boundaries along with a pressure reference in some point in the camera frame. Since we assume that the particle does not influence the fluid velocities in the network outside the camera, if we are able to determine the velocity profiles in the network, then the knowledge of the translations and rotation(s) of the camera gives the positions of the boundaries in the network and the velocities to impose. To determine those velocity profiles (and pressures), we can either use analytical calculation if possible (for example in the case of a long tube in Poiseuille's regime) or a preliminary numerical simulation of the fluid in the network without any particle (pure computational fluid dynamics simulations).

A part of the camera frame can get out of the fluid network when the particle comes too close to the network walls. In the part outside the network, the fluid does not physically exist, but its velocity can be extended mathematically with a null velocity. This operation leads to fluid velocities that are continuous in the whole camera frame and that are consistent with wall boundary conditions in the network. This is integrated in the fluid equations thanks to a penalization method \cite{Peskin,Cottet}. Note that we assume that the camera frame does not get out of the network near an inlet or an outlet. If this happens, then the extension of velocity is still feasible but more complex, all the more if the camera gets simultaneously out of a wall boundary.

\subsection{General equations}

In this section we will describe the mathematical equations involved in the camera method. Since this method is inspired from the standard ALE method, we will recall first the equations used for that method.

\subsubsection{Standard ALE equations}

\begin{figure}[h!]
\centering
\begin{psfrags}
\psfrag{Gin0}{$\Gamma_{in,0}$}
\psfrag{Gout0}{$\Gamma_{out,0}$}
\psfrag{Gw0}{$\Gamma_{w,0}$}
\psfrag{S0}{\small{$S_0$}}
\psfrag{F0}{$F_0$}
\psfrag{reference frame}{reference frame $\Omega_0$}
\psfrag{coordinates (x)}{coordinates $(x)$}
\psfrag{solid equations on S0}{solid equations on $S_0$}
\psfrag{Gint}{$\Gamma_{in,t}$}
\psfrag{Goutt}{$\Gamma_{out,t}$}
\psfrag{Gwt}{$\Gamma_{w,t}$}
\psfrag{St}{\small{$S_t$}}
\psfrag{Ft}{$F_t$}
\psfrag{y=phi(x,t)}{$y=\phi(x,t)$}
\psfrag{deformed frame}{deformed frame $\Omega_t$}
\psfrag{coordinates (y)}{coordinates $(y)$}
\psfrag{fluid equations on Ft}{fluid equations on $F_t$}
\includegraphics[height=5cm]{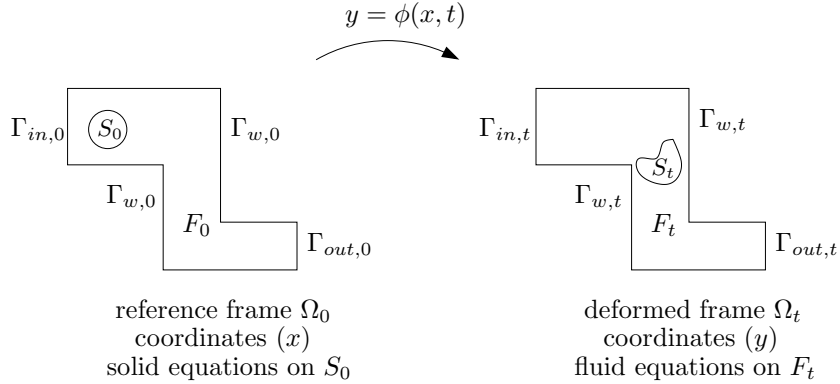}
\end{psfrags}
\caption{2D example of the initial domains of the physical problem (left) that corresponds to the reference frame and of the deformed frame (right). particle equations are written in the reference frame while fluid equations are written in the deformed frame. The application $x \rightarrow y=\phi(x,t)$ transforms the coordinates of the reference frame into the coordinates of the deformed frame.}
\label{ALEdom}
\end{figure}

As shown on figure \ref{ALEdom}, we will consider a fluid network $\Omega_0 \subset \mathbb{R}^N$ ($N=2$ or $3$)) which has inlet(s) $\Gamma_{in,0} \subset \partial \Omega_0$, outlet(s) $\Gamma_{out,0} \subset \partial \Omega_0$) and walls ($\Gamma_{w,0} \subset \partial \Omega_0$) such that $\partial \Omega_0 = \Gamma_{in,0} \cup \Gamma_{w,0} \cup \Gamma_{out,0}$. In this network, a particle fills initially the subdomain $S_0$ of $\Omega_0$. Consequently, the initial fluid domain is $F_0=\Omega_0 \backslash S_0$. This geometry corresponds to the reference frame where the equations of the particle mechanics are defined. We call $(x)$ the coordinates in the reference frame.

The deformed frame at time $t$ corresponds to the physical frame where the particle has moved and is deformed. The fluid domain in the deformed frame is the one in which the fluid physically spans at time $t$ and thus where the Navier-Stokes equations are defined. The deformed frame is assumed to be the image by a smooth derivable and invertible (typically a $\mathcal{C}^1$ diffeomorphism) application $\phi$ of the reference frame, we call $(y)$ the coordinates in the deformed frame and $y=\phi(x,t)$. For each subset $X_0$ of $\mathbb{R}^N$ defined in the reference frame, we define $X_t$ its image by $\phi$ in the deformed frame i.e. $X_t=\phi(X_0,t)$, see figure \ref{ALEdom}.

The solid equations are written in the reference frame of the solid, i.e. on $S_0$, (Lagrangian equations) while the fluid equations are written in the deformed frame where the solid has moved and is deformed (Eulerian equations). Consequently, we need to determine the transformation $\phi$ that transforms the coordinates of the reference frame ($x$) into the coordinates of the deformed frame ($y=\phi(x,t)$). 

From now on, the fluid velocity is represented by $v(y,t)$, the fluid pressure $p(y,t)$ and the structure displacement $u(x,t)$. The fluid and solid densities are respectively denoted $\rho_f$ and $\rho_s$ and the fluid viscosity is denoted $\eta_f$. The fluid constraints tensor is $\sigma_f(v,p) = -p I + \eta_f \left( \nabla u + \phantom{}^t \nabla u \right)$ and the solid tensor constraint is $\sigma_s(u)$.  The transformation $\phi$ is known in the solid $S_0$ since for $x \in S_0$, $\phi(x,t)= x + u(x,t)$. It is however necessary to define the transformation $\phi$ on the reference fluid domain $F_0$, in order to be able to write the fluid equations. A way to define $\phi$ on $F_0$ is to use the Laplace equations:

\begin{equation}
\label{IFStrans}
\left\{
\begin{array}{ll}
\bigtriangleup \phi = 0 & \text{on $F_0$}\\
\phi(x,t) = x + u(x,t) & \text{on $\partial S_0$}\\
\phi(x,t) = x & \text{on $\partial \Omega_0$}
\end{array}
\right.
\end{equation}

Note that we assume that the boundary $\partial \Omega_0$ is unmoving, thus the transformation is the identity on $\partial \Omega_0$.

Incompressible fluid-structure interaction is governed by the mechanics equations for the structure in the reference frame ($x$) and the equations of Navier-Stokes for the fluid in the deformed frame ($y=\phi(x,t)$).

The mechanics equations are:

\begin{equation}
\label{IFSsolid}
\left\{
\begin{array}{ll}
\rho_S \frac{\partial^2 u}{\partial t^2} - div\left(\sigma_s(u)\right) = 0& \text{on $S_0$}\\
\sigma_s(u).n_s = - \sigma_f(v,p).n_f& \text{on $\partial S_0$} 
\end{array}
\right.
\end{equation}

The constraints on the boundary of the solid are equal to the fluid constraints on the solid boundary {\bf brought back to the reference frame} thanks to the application $\phi$. The minus sign is due to the orientation of the normals $n_s$ and $n_f$ which are defined as outwards normals in their respective subset.

The Navier-Stokes equations are:

\begin{equation}
\label{IFSfluid}
\left\{ 
\begin{array}{ll}
\rho_f \frac{\partial v}{\partial t} + \rho_f \left(v.\nabla\right)v - 
div\left( \sigma_f(v,p) \right)
= 0& \text{on $F_t=\phi(F_0)$}\\
div(v) = 0& \text{on $F_t=\phi(F_0)$}\\
v(y) = \frac{\partial u}{\partial t}\left( \phi^{-1}(y,t) \right)& \text{on $\partial S_t = \phi(\partial S_0)$}\\
v(y) = 0 & \text{on $\Gamma_{w,t}$}\\
+\text{inlet and outlet boundary conditions on $\Gamma_{in,t}$ and $\Gamma_{out,t}$}&
\end{array} 
\right.
\end{equation}

Due to fluid viscosity, the fluid is sticking on the particle and fluid velocity at particle boundary is equal to the particle velocity (second to last equality).

In the next section, we use the equations of the standard ALE method to define the equations of the camera method.

\subsubsection{Camera method}

We consider a neighbourhood $C$ of $S_0$, typically a sphere centred on the barycentre of $S_0$. $C$ is the frame of the camera. We do now the hypothesis that the particle $S_0$ does not influence the fluid outside of $C$. We will discuss the meaning of this hypothesis later in this paper. We will assume that the barycentre of the particle $S_0$ is initially at the origin of the coordinates frame (i.e. $\int_{S_0} x dx = 0$), because it simplifies the writing of equations (see below). This is not restrictive because a simple translation of the domains makes this condition true.

On the contrary of the standard ALE method, the camera method works by solving only a subset of the full problem. The fluid-structure interaction problem is restricted to the neighbourhood $C$ of the particle. Moreover, the mesh of the neighbourhood $C$ of the particle moves and rotates with the particle in order to avoid most of the mesh deformation. Two points need to be addressed to solve that new problem. First, fluid velocities (and fluid pressure) are not known on the boundaries of $C$, so we will use an a priori estimation of these quantities in absence of the particle. Secondly, in the general case, the neighbourhood $C$ does not fit the network geometry and some parts of $C$ can be outside the network where no fluid physically exists (see an example on figure \ref{bif2d} (right) or figure \ref{camfull}). This second point is solved by making the fluid virtually span outside $\Omega_t$ and by setting its velocity there to zero with a penalization method (also called immersed boundary method) \cite{Peskin, Cottet}.\\

{\bf Decomposition of the structure displacement.} This method is based on the unique decomposition of the solid displacement $u(x,t)$ under the hypothesis that the gravity center of the solid $S_0$ is at the origin of the reference frame ($\int_{S_0} x dx = 0$) \cite{Salomon}:

\begin{equation}
x+u(x,t) = \tau(t) + R_{\theta(t)} \left( x+d(x,t) \right) \text{ ($= \phi(x,t)$ on $S_0$)}
\label{decomp}
\end{equation}

where :

\begin{itemize}
\item $(x,t) \rightarrow d(x,t)$ is an "elementary" displacement without any translation or rotation, i.e.:

\begin{equation}
\label{SALdechyp}
\begin{array}{ll}
\int_{S_0} d(x,t) dx = 0 &\text{ no translation }\\
\int_{S_0} x \wedge d(x,t) dx = 0 &\text{ no rotation }\\
\end{array}
\end{equation}

\item $t \rightarrow \tau(t)$ is a vector representing the global translation of $S_0$ at time $t$.

\item $R_{\theta(t)}$ is an invertible matrix that reflects  the rotation(s) of $S_0$ at time $t$. The definition of $R_{\theta(t)}$ depends on the dimension of the space, see the next sections for more details. We will denote $R_{-\theta(t)}$ the inverse of the matrix $R_{\theta(t)}$.

\end{itemize}

Our goal is not to calculate the function $d$ but to use its existence to determine the translation $\tau(t)$ and the rotations $\theta(t)$. Once these quantities are determined, they are used to make the mesh translate and rotate with the solid, thus avoiding most of mesh deformations. Whatever the dimension of the space, the translation is given by

\begin{equation}
\tau(t) = \frac{\int_{S_0} u(x,t) dx}{\int_{S_0} dx}
\end{equation}

Finally, the fluid structure interaction equations can be rewritten using these new information. The major changes will affect the transformation and the representation of the fluid.

The equations that determine the transformation $\phi$ become:

\begin{equation}
\label{CAMtrans}
\left\{
\begin{array}{ll}
\bigtriangleup \phi = 0 & \text{on $F_0$}\\
\phi(x,t) = x + u(x,t) & \text{on $\partial S_0$}\\
\phi(x,t) = R_{\theta(t)}x + \tau(t) & \text{on $\partial C$}
\end{array}
\right.
\end{equation}

This defines the coordinate transformation we used in our work. It lets the mesh centered on the particle $S$ and makes it rotate with the particle. There is an alternative way to define the transformation $\phi$ by taking for boundary conditions on $\partial S_0$ and $\partial C$: $\phi(x,t) = x + u(x,t) - \tau(t)$ on $\partial S_0$, and $\phi(x,t) = R_{\theta(t)}x$ on $\partial C$. This alternative transformation makes the visualisation process easier, since the mesh remains centered on the origin at each time step.\\

{\bf Fluid domain and boundary conditions.} Since we limit the fluid domain to the frame of the camera that corresponds to a neighbourhood of the structure $S$, we need to be able to impose zero velocity to the fluid if a part of the camera frame gets out of $\Omega_t$. This is achieved thanks to a penalisation method. Using the function $\chi(y)$ that is equal to $0$ in $\Omega_t$ and to $1$ in $\mathbb{R}^n \backslash \Omega_t$, then the fluid equations become

\begin{equation}
\label{CAMfluid}
\left\{ 
\begin{array}{ll}
\rho_f \frac{\partial v}{\partial t} + \rho_f \left(v.\nabla\right)v - 
div\left( \sigma_f(u,p) \right) + \rho_f \frac{\chi}{\epsilon} v
= 0& \text{on $F_t=\phi(F_0)$}\\
div(v) = 0& \text{on $F_t=\phi(F_0)$}\\
v = \frac{\partial u}{\partial t}& \text{on $\partial S_t = \phi(\partial S_0)$}\\
v = v_0 & \text{on $\partial C$}\\
p = p_0 & \text{on a point $P$ (reference pressure)}
\end{array} 
\right.
\end{equation}

The difficulty arises in the determination of the velocities $v_0$ on the boundary of the camera ($\partial C$). Since we hypothesized that the particle does not influence the fluid outside the camera, $v_0$ can be determined as the result of an analytical calculation or a preliminary numerical simulation of the fluid in the network without the particle. The reference pressure $p=p_0$ on an arbitrary point $P$ in the camera frame is required for pressure uniqueness. The point $P$ moves with the camera and consequently, the pressure reference is time-dependent. However, this does not influence the pressure drops nor the fluid velocities which are correctly computed, since the time-dependent reference pressure disappears under the spatial gradient that operates on the pressure in the fluid equation.

\section{Camera method set up and validation through examples}

The set up of the camera method depends on the space dimension. The translation vector is easily computed whatever the number of dimensions. However the definition of rotations depends on the number of dimensions in the space. In this section, we describe how to set up the camera method in a 2D axi-symmetric space, in a 2D space and in a 3D space. For each case, we explain how the number of dimensions affects the camera method and for each case, we validate with an example from the literature. Furthermore, a 2D axi-symmetric space is well adapted to the tracking of a periodic train of particles along a straight channel. Thus, we outline how the camera method can be used in such a context with a model of red blood cells in a capillary.

The numerical simulations performed in this work were done with the commercial finite elements package Comsol Multiphysics, with the linear solver PARDISO. The computations are performed on eight cores of a bi-processors workstation (two Xeon E5645) with 32 GBytes of memory. The most memory consuming computation (standard ALE method, see below) needs about 4 GBytes to run.

\subsection{2D axi-symmetry}

The axis of axi-symmetry is referred to as the $(z)$ axis while the coordinates along the radius is referred to as the $(r)$ axis. In the following sections, we assume that all quantities are independent on the phase, thus we write vector coordinates in the form $v=(v_r,v_z)$. 

\subsubsection{Specificity of camera method in 2D-axi}

Both the particle and the network are assumed axi-symmetric. The model is able to represent a particle moving along the axis of a unidirectional channel whose section is circular. The radius of the channel is not necessarily constant and can depend on $z$ coordinate. The camera is bound to move along the axis of the channel, and no camera or particle rotation is possible, i.e. $\theta(t) = 0$ for each time $t$:

$$
R_{\theta(t)} = I
$$

The movement of the barycentre of the particle is limited to a translation along the $z$-axis, consequently $\tau(t)$ has only one non zero component on the $z$ coordinate: $\tau(t) = (0,\tau_z(t))$.

\begin{figure}[h!]
\centering

\begin{psfrags}
\psfrag{Wall_}{fluid penalization ($\chi=1$ implies $v \sim 0$)}
\psfrag{CintF1_}{$C \cap F_t$}
\psfrag{CintF2_}{$C_t \cap F_t = C_t$}
\psfrag{S_}{$S_t$}
\psfrag{I_}{$I_t$}
\psfrag{E_}{$O_t$}
\psfrag{W_}{$W_t$}
\psfrag{C1_}[c]{camera frame $C$}
\psfrag{C2_}[c]{camera frames $C_0$ and $C_t$}
\psfrag{boundarydisp1_}{frame upper boundary}
\psfrag{boundarydisp2_}{sticks to channel wall}
\psfrag{initialframeshape_}{initial camera frame ($C_0$)}
\psfrag{frameshape_}{camera frame at time $t$ ($C_t$)}
\includegraphics[height=4cm]{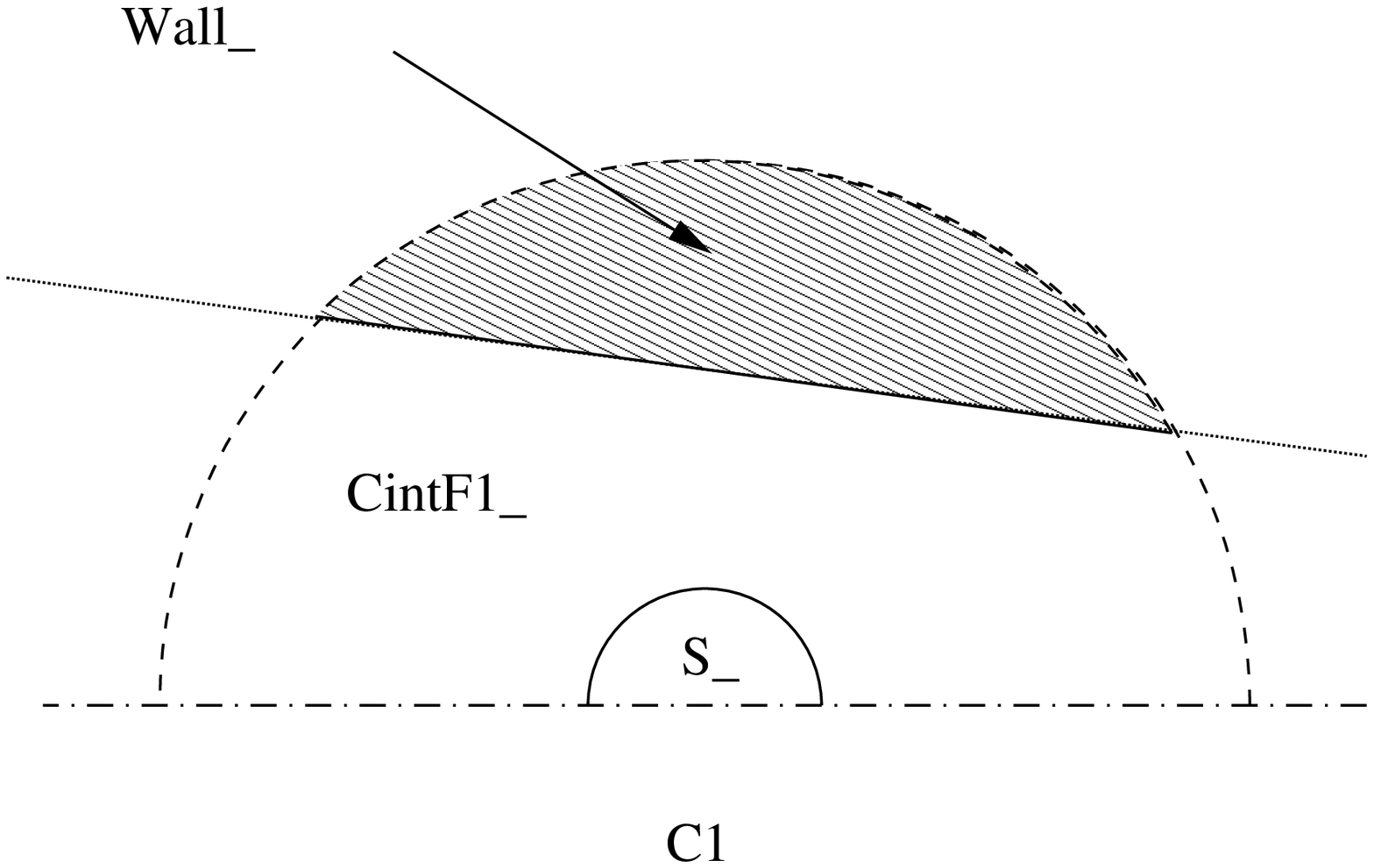}
\hspace{1cm}
\includegraphics[height=4cm]{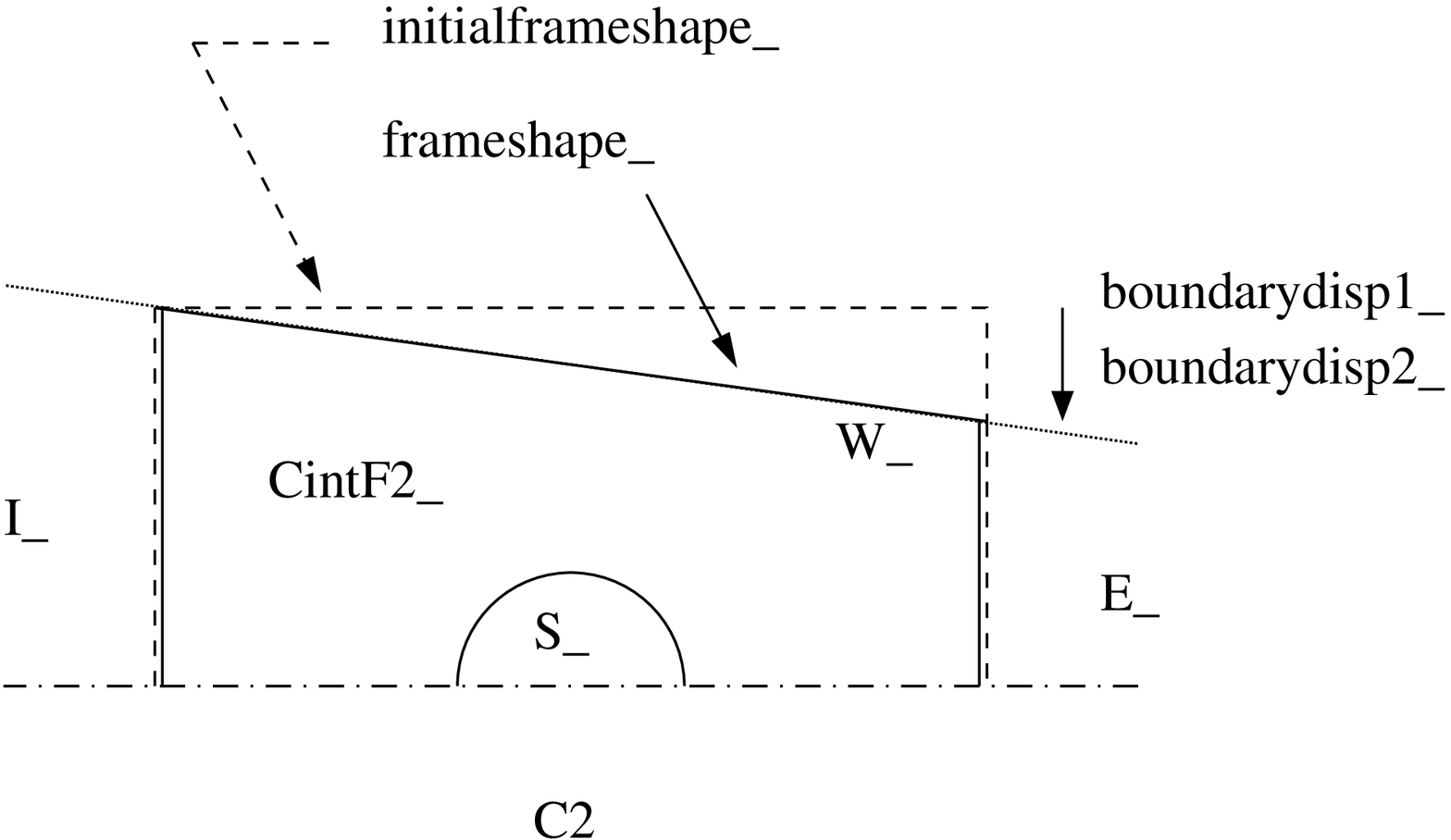}
\end{psfrags}
\caption{Two ways to define camera frames in 2D axi-symmetry. Left: ``classic'' camera frame, penalization is used to nullify fluid velocity in the intersection between channels walls and camera frame. Right: the upper border of the camera frame moves such that it fits the wall geometry of the channel at any time, the frame of the camera never crosses the walls of the channel.}
\label{camframe2Daxi}
\end{figure}

There are two possible ways to define the camera frame when working in 2D axi-symmetry. The first way is to work with a fixed frame and to use penalization to nullify the fluid velocity where the wall crosses the camera frame (see left part of figure \ref{camframe2Daxi}). So the general equations of the previous section apply. The other way is to use the specificity of the 2D axi-symmetry and to make the ``upper'' border of the camera frame fit the wall geometry (see right part of figure \ref{camframe2Daxi}). This second method is well adapted to axi-symmetry. Although it reduces slightly the mesh quality, it avoids the use of penalization that can sometimes lead to convergence problems. The equations of the transformation $\phi$ are slightly different than equations \ref{CAMtrans} and they need a parametrization of the channel radius along the axis: $R: \ z \longrightarrow R(z)$. The transformation $\phi$ is then solution of the equations ($x=(r,z)$):

\begin{equation}
\label{CAMtrans2Daxi}
\left\{
\begin{array}{ll}
\bigtriangleup \phi = 0 & \text{on $F_0$}\\
\phi(x,t) = 
\left(
r + u_r(x,t),
z + u_z(x,t)
\right) & \text{on $\partial S_0$}\\
\phi(x,t) = 
\left(
R(z + \tau_z(t)),
z + \tau_z(t)
\right) & \text{on $W_0$}\\
\phi(x,t).n=0 & \text{on $I_0 \cup O_t$}\\
\end{array}
\right.
\end{equation}

The fluid boundary conditions need also to be slightly adapted: on the ``upper'' boundary of the camera frame ($W_t$), we use no-slip boundary conditions. On inlet $I_t$ and outlet $O_t$ fluid velocity conditions (Dirichlet) are imposed from an a priori estimation as before (numeric or analytic). If the Reynolds number is low enough and if the width of the camera is large enough, one can use Poiseuille profiles as approximations.

In the following, we will systematically use the alternative method (camera border fitting channel wall) when dealing with 2D axi-symmetry problems. In particular this alternative method can be slightly adapted to model infinite trains of particles in straight channels, such as red blood cells in capillaries. It is achieved by adding a periodicity condition on the boundary of the camera frame and a Lagrange multiplier in the equations, see section \ref{RBCper}.

\subsubsection{2D-axi validation: deformation of a vesicle in a narrowing channel}
\label{risso2Daxi}

A vesicle consists in a thin membrane enclosing an inner fluid \cite{Quequ}, vesicles are deformable objects able to undergo strong deformation. Red blood cells are natural vesicles that carry oxygen in blood \cite{Weibel}, bio-artificial vesicles also exists and can be used to carry medicine to a precise location. When such an object is motioned by an outer fluid in a narrow channel, such as the capillaries for red blood cells, then its stationary shape is a "parachute" shape \cite{Fung}. The characteristics of that shape depends on the different physical parameter involved in the outer and inner fluid and in the membrane. Experimental and numerical works \cite{Risso, Barthes1, Diaz, Quequ} have been made to study the behaviour of a vesicle in a narrowing channel. The numerical simulations made in \cite{Risso, Quequ} make a good benchmark to validate the camera method in 2D axi-symmetry. In these works, the vesicles are droplets of salt water surrounded by a thin polymeric membrane whose thickness is about $30 \ \mu m$. Their size is millimetric.

We assume that the vesicle is initially a sphere of radius $a$ and that it is motioned through a narrow channel of radius $R$. The behaviour of the vesicle is determined by \cite{Risso, Quequ}: 1/ the capillary number $Ca=\frac{\eta_{ext} U}{K}$, where $\eta_{ext}$ is the viscosity of the outer fluid, $U$ the bulk velocity of the fluid and $K$ the membrane area dilatation modulus ($N.m^{-1}$); 2/ the ratio $a/R$ between the vesicle radius and the channel radius.

In our simulations, the membrane is a full 2D axi-symmetric material, whose thickness is that of the experimental object, $30 \ \mu m$. We assume the material to be elastic and that it undergoes large deformation. Its Young modulus is $E$ and its Poisson's ratio is $\nu$. In the numerical works proposed in \cite{Risso, Barthes1, Quequ, Diaz}, the authors used a boundary integral method with an infinitely thin membrane. The parameters $E$ and $\nu$ used in our simulations were chosen such that the material properties fits the infinitely thin membrane models used in \cite{Risso, Quequ}:

$$
\nu = \frac{K-\mu}{K+\mu} \ \ \text{ and } \ \ E = \frac{4 K \mu}{h (K+\mu)}
$$

\begin{figure}[h!]
\centering
\begin{psfrags}
\psfrag{rv_}{$1.58 \ mm$}
\psfrag{R_}{$8 \ mm$}
\psfrag{r_}{$2 \ mm$}
\psfrag{L_}{$10 \ cm$}
\includegraphics[height=4cm]{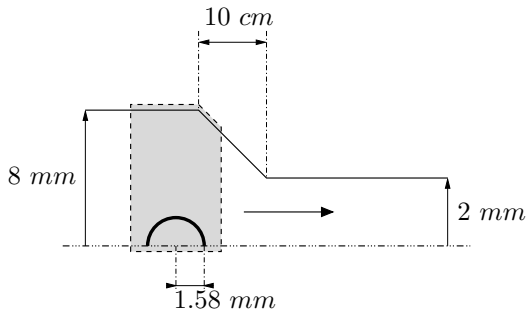}
\end{psfrags}
\caption{Numerical model of experiments from \cite{Risso}: 2D axi-symmetric geometry of the channel, vesicle and camera frame (grayed box). The arrow reflects the orientation of the fluid flow and the direction in which the vesicle and the camera are moving.}
\label{channelrisso}
\end{figure}

The vesicle is injected into a wide channel whose radius decreases to reach a radius of $R=2 \ mm$ close to the radius of the vesicle which needs to deform to enter the constriction. We model the channel with the geometry shown on figure \ref{channelrisso}. First the vesicle is accelerated in the wide section of the channel, then it enters the constriction. 

The upper and lower boundary of the camera frame coincide respectively to the channel wall and to the symmetry axis. The left and right borders are fluid filled sections of the channel, see figure \ref{channelrisso}. The fluid boundary conditions on the left and right borders are assumed to be parabolic velocity profiles. This last hypothesis is an approximation, however the Reynolds number value is low and the camera frame width is large relatively to the distance needed for the fluid to be fully developed. Thus the fluid is fully developed far before it reaches the neighbourhood of the vesicle.

\begin{figure}[h!]
\centering
\includegraphics[height=5cm]{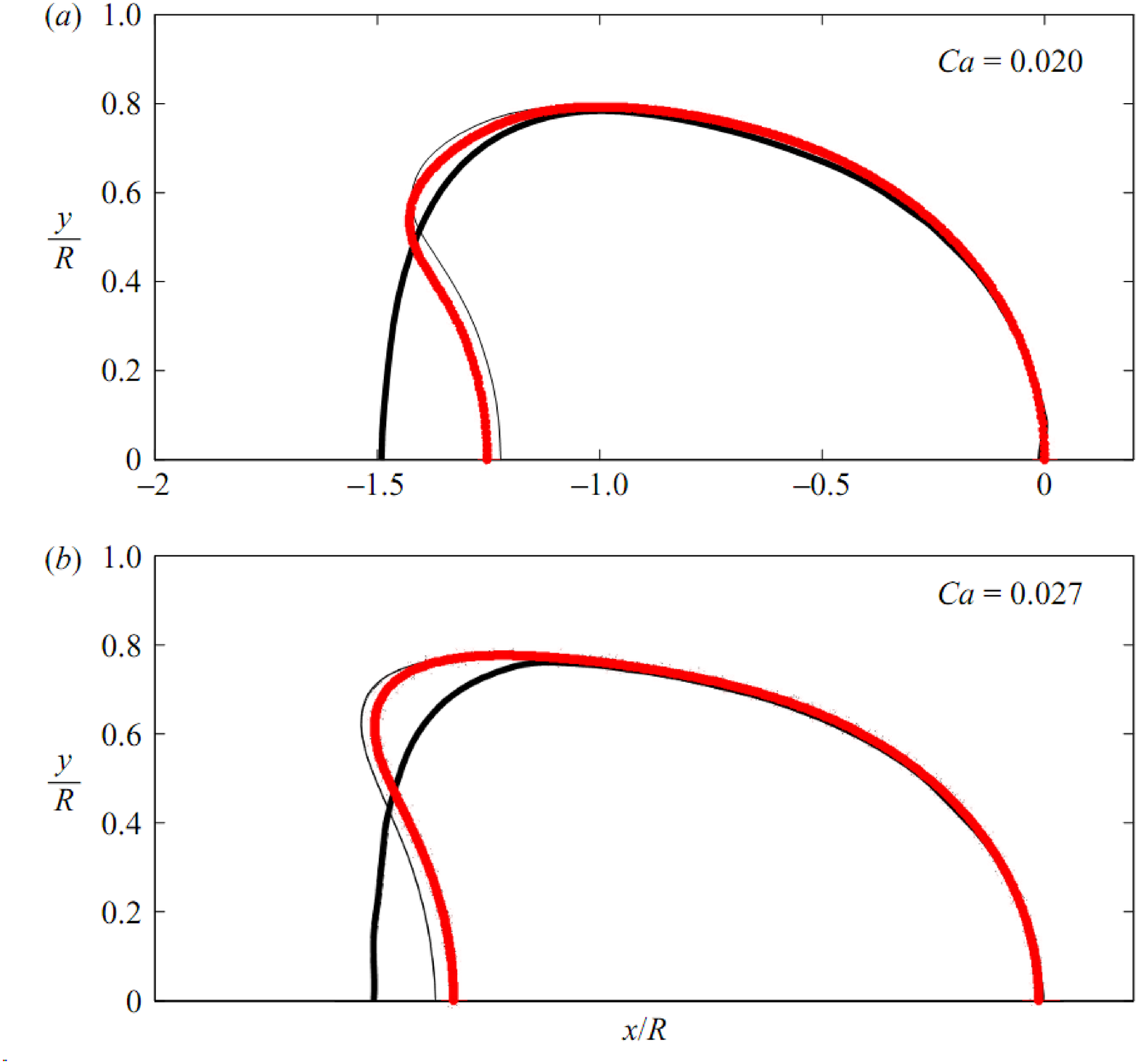}
\caption{Shapes of the vesicles computed with camera method (red thick lines) for $a/R = 0.78$. Up: $Ca=0.02$, $mu/K \sim 1$; Down: $Ca=0.027$, $\mu/K \sim 1/3$.  The thick black lines corresponds to the experiments made by Risso et al \cite{Risso}, the thin black lines corresponds to the results of simulations made by Diaz and Barth\`es-Biesel \cite{Diaz} with infinitely thin membrane model. The coordinates are normalized with the radius of the constricted section of the channel ($R=2 \ mm$).}
\label{risso_barthes}
\end{figure}

We computed the deformed shape of a vesicle whose $a/R$ ratio is $0.78$ and whose membrane area dilatation modulus is $K=1.30 \ N.m^{-1}$. Two cases were simulated: $\mu / K = 1$, $Ca = 0.02$ and $\mu / K = 1/3$, $Ca=0.027$. Our results are compared with the experimental shapes and numerical simulations from \cite{Risso, Diaz} on figure \ref{risso_barthes}. Volume variations of the vesicle during our computations are less than $0.02\%$, which is compatible with the incompressibility assumption for the inner fluid. The numerical simulations from \cite{Diaz} and our simulations give close results with a slight difference at the rear of the vesicle, which is probably a consequence of the different models used for the membranes. As discussed in \cite{Risso}, the numerical shapes match well the front and intermediate parts of the experimental vesicle but not the rear.

\subsubsection{2D axi extension: periodic train of red blood cells in a capillary}
\label{RBCper}

\begin{figure}[h!]
\centering
\begin{psfrags}
\psfrag{Ct_}[c]{$S_t$}
\psfrag{I_}{$I$}
\psfrag{O_}{$O$}
\psfrag{W_}{$W$}
\psfrag{Ft_}[c]{$F_t$}
\psfrag{cam}{periodic camera frame}
\includegraphics[height=3cm]{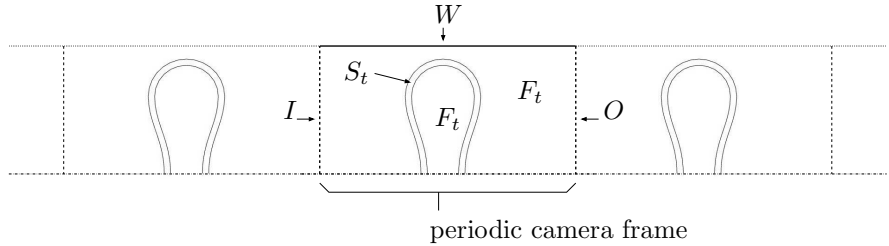}
\end{psfrags}
\caption{Frame of the camera for the axi-symmetric model of a periodic train of red blood cells. The dashed-dotted line represents the axis of axi-symmetry. Periodic boundary conditions are applied on $I$ and $O$ boundaries: velocities and normal constraints are equal on $I$ and $O$.}
\label{RBCaxi}
\end{figure}

In this section, we show with an example how to use the camera method to model a train of particles in a straight channel. We model here a train of red blood cells going through an idealized capillary. The red blood cells are modeled as discoid vesicles. The vesicle diameter is $7.5 \ \mu m$ and the vesicle thickness ranges from $1 \ \mu m$ on their center to $2 \ \mu m$ near their periphery \cite{Weibel}. Half sections of the vesicle are plotted on figure \ref{RBCaxi}. The frame of the camera contains one vesicle on which it is centered. The frame is rectangular and unchanging during the simulation since the capillary is assumed to have a constant diameter of $8 \ \mu m$. The upper boundary corresponds to the wall of the capillary and the lower boundary to the axi-symmetry axis. The left and right boundaries correspond to capillary sections that moves with the vesicles.
As for the previous example, it is not necessary to use fluid penalization since the top border of the camera always coincides with the wall of the capillary.
A scheme of the camera frame is plotted on figure \ref{RBCaxi}. 

The capillary consists in successive copies of the camera frame that are connected by their left and right boundaries ($I$ and $O$). The number of red blood cells volumetric fraction can be easily modulated by changing the width of the camera frame. Mathematically, this succession of "cells'' can be represented with periodic boundary conditions for the fluid on $I$ and $O$. The fluid equations are (see for example \cite{Amrouche}):

\begin{equation}
\label{fluidper}
\left\{ 
\begin{array}{ll}
\rho_f \frac{\partial v}{\partial t} + \rho_f \left(v.\nabla\right)v - 
div\left( \sigma_f(u,p) \right)
= 0& \text{on $F_t$}\\
div(v) = 0& \text{on $F_t$}\\
v = \frac{\partial u}{\partial t}& \text{on $\partial S_t$}\\
v = 0 & \text{on $W$}\\
p = 0 & \text{on an arbitrary point (reference pressure)}\\
v|_I = v|_O&\\
\nabla v . n |_I = - \nabla v . n |_O&\\
\int_I v.n ds = Q&\\
\end{array} 
\right.
\end{equation}

The third to last equality in equations (\ref{fluidper}) states that fluid velocity profiles are identical on $I$ and $O$. The last but one equality states that fluid viscous constraints are identical on $I$ and $O$. Both equalities define the periodic boundary conditions for the fluid, but they are not sufficient to close the problem, since they do not define a flow rate in the capillary. The last equality is thus needed, it fixes the flow rate to a given value $Q$. The particles velocity in the capillary can be tuned by changing the flow rate value $Q$. In the weak formulation of equations (\ref{fluidper}), the constraint on the flow rate brings a supplementary term which is a Lagrange multiplier times the differential of the constraint relatively to the fluid velocity $v$.

The membrane of the red blood cell consists in a bilipidic layer stacked up with a spectrin mesh. Thick hyperelastic material such as Yeoh's model fits well the behaviour of red blood cells membrane \cite{Yeoh, Mills, MauroyRBC}. Yeoh's energy of deformation is:

$$
W_s = \frac{G_0}{2} \left( \lambda_1^2+\lambda_2^2+\lambda_3^2 -3\right) + C_{yeoh} \left( \lambda_1^2+\lambda_2^2+\lambda_3^2 -3\right)^3 + \frac{k_0}{2} \left( \lambda_1^2 \lambda_2^2 \lambda_3^2 - 1 \right)^2
$$

where $G_0$ is the membrane shear modulus, $k_0$ the membrane elastic modulus and $C_{yeoh}$ the Yeoh's constant. The term $\lambda_1^2+\lambda_2^2+\lambda_3^2$ is the first invariant (trace) of the right Cauchy-Green tensor which is equal to identity when no deformation occurs. The strain energy is thus governed by its first term at smaller deformations. On the contrary for large deformations, the second term is dominant and the strain energy grows more rapidly (power three versus power one). The last term corresponds to the volume change of the membrane which is given by the third invariant (determinant) of the right Cauchy-Green tensor $\lambda_1^2 \lambda_2^2 \lambda_3^2$. A large bulk modulus $k$ ensures that the membrane volume change remains small. To model the red blood cells, we made the same hypothesis than \cite{MauroyRBC} and used a membrane which is ten times thicker than the real membrane (i.e. $100 \ nm$ instead of around $10 \ nm$) and we rescaled the membrane mechanical parameters to reflect its real thickness. The data used are from \cite{Mills, MauroyRBC}: $G_0=56.5 \ Pa$, $k_0=100 \times G_0$ and $C_{yeoh}=G_0/30$.

\begin{figure}[h!]
\centering
\begin{psfrags}
\psfrag{Hc_11}[c]{$Hc=0.11$}
\psfrag{Hc_18}[c]{$Hc=0.18$}
\psfrag{Hc_25}[c]{$Hc=0.25$}
\includegraphics[width=12cm]{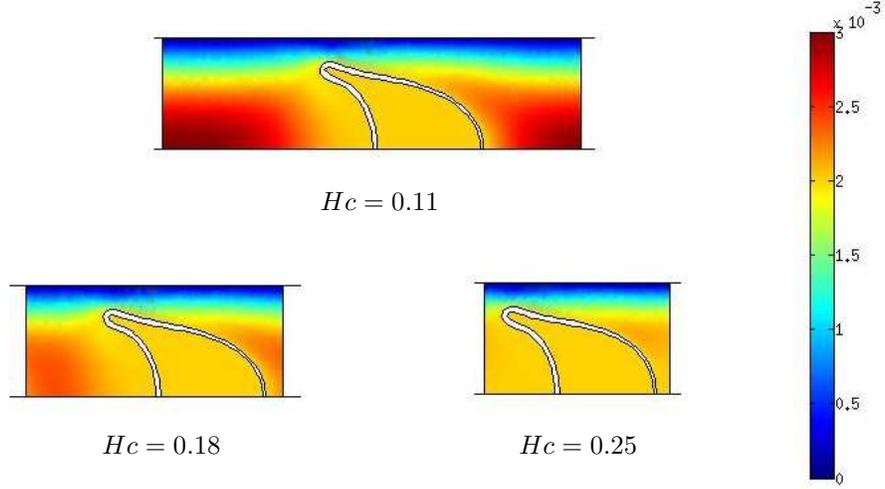}
\end{psfrags}
\caption{Results of 2D axi-symmetric camera simulations: stationary shapes of models of red blood cells for different value of their volumetric fraction $Hc$ in the channel. The color represents the amplitudes of fluid velocities (in plasma or vesicle cytosol, in $m.s^{-1}$).}
\label{RBCres}
\end{figure}

\begin{figure}[h!]
\centering
\includegraphics[height=5cm]{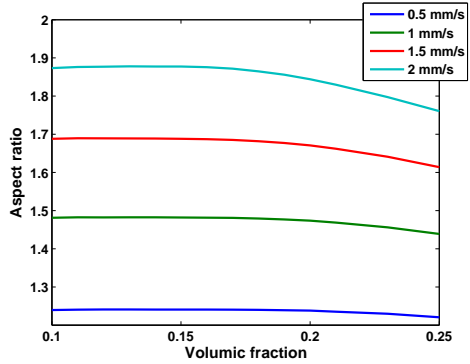}
\caption{Discoid vesicle aspect ratio (length over diameter) versus volumetric fraction of vesicles in the channel for different vesicle velocities.}
\label{RBCHC}
\end{figure}

We tested the role of the volumetric fraction of red blood cell-like vesicles in the channel on the shape of the vesicles for stationary regime, see figure \ref{RBCres}. In stationary regime, the vesicles takes the shape of a parachute. The stationary parachute shape changes with vesicles volumetric fraction $Hc$ because of disturbance due to the neighbouring cells. The aspect ratios of the discoid vesicles versus their volumetric fraction is plotted on figure \ref{RBCHC}.

\subsection{2D}
\label{2D}

In this section, we describe how to implement the camera method for bi-dimensional particle and fluid. We use the coordinates $x=(x_1, x_2)$ for the reference frame and $y=(y_1,y_2)$ for the deformed frame.

\subsubsection{Specificity of camera method in 2D}

In 2D, the particle is only able to rotate around an axis perpendicular to the spanning plan of the particle and fluid. Thus, the rotation part $R_{theta(t)}$ in the decomposition (\ref{decomp}) is a 2D rotation matrix in that plane:

$$
R_{\theta(t)} = \left(
\begin{array}{cc}
\cos(\theta(t)) & -\sin(\theta(t)) \\
\sin(\theta(t)) & \cos(\theta(t))
\end{array}
\right)
$$

The number $\theta(t)$ is the rotation angle of the particle at time $t$.  Generally, the rotation angle reference is chosen for $t=0$, i.e. $\theta(0)=0$. The angle $\theta(t)$ is computed from the rotation constraint on the elementary displacement $d(x,t)$, see equations (\ref{SALdechyp}). Thus, $\theta(t)$ is computed by solving the equation: $\int_{S_0} x \ \wedge \ d(x,t) \ dx = \int_{S_0} x \ \wedge R_{-\theta(t)}\left( x + u(x,t) - \tau(t) \right) \ dx = 0$. The angle of rotation $\theta(t)$ can then be computed explicitly as a function of the particle displacements $u$:

$$
\tan\left(\theta(t)\right) = \frac{\int_{S_0} x \wedge u(x,t)dx}{\int_{S_0} x.\left(x+u(x,t)\right)}
$$

In order to have a unique solution, $\theta(t)$ is computed from this last formula using a function $atan2$.

Ideally, the 2D camera frame is a disk centered on the particle, such as the frame on figure \ref{2Dpartshear}. A disk-shaped frame remains still whatever the rotation, which makes post-processing easier to interpret. However, the disk shape is not compulsory and any other shape is possible as soon as it encloses the particle and is wide enough to be able to neglect the particle influence outside of the camera frame.

\subsubsection{2D validation: a deformable particle in shear flow}

A circular elastic particle in a shear flow deforms in an ellipse which rotates around its gravity center. This phenomena is well known, see for example \cite{GaoHu, Gold}, and is often used to validate numerical work like in \cite{Shizhi}.

\begin{figure}[h!]
\centering
\begin{psfrags}
\psfrag{vit_}{$v(y)=\left( \begin{array}{c} \dot{\gamma} y_2 \\ 0 \end{array} \right)$}
\psfrag{x_}{$y_1$}
\psfrag{y_}{$y_2$}
\includegraphics[width=6cm]{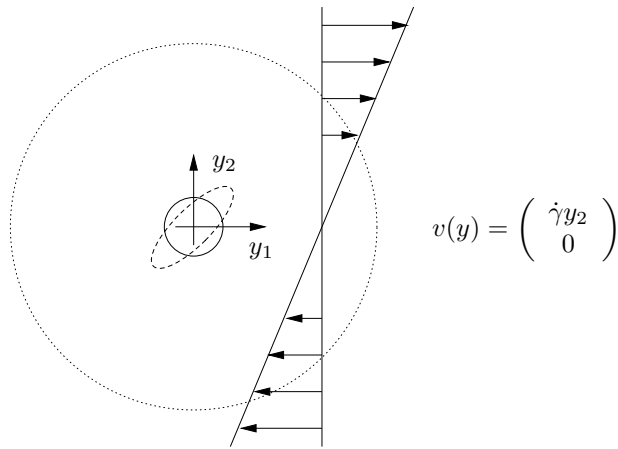}
\end{psfrags}
\caption{The elastic particle stands in a shear flow. The camera frame is represented by the dotted circle enclosing the particle. The particle deforms in an ellipsoid shape \cite{GaoHu}. }
\label{2Dpartshear}
\end{figure}

The particle at rest is a disk whose diameter is $d_p=1 \ mm$. The particle material is a Neo-Hookean material whose density is $1000 \ kg.m^{-3}$ and whose shear modulus is $G=1 \ Pa$. The fluid enclosing the particle is water (viscosity $\eta_{ext} = 10^{-3} \ Pa.s$, density $\rho_{ext}=1000 \ kg.m^{-3}$) and it spans infinitely in all directions. The shear flow is defined from the shear rate $\dot{\gamma}$ which is constant in space. The fluid velocity in the absence of the particle is then known everywhere in space:

\begin{equation}
v(y)=
\left(
\begin{array}{c}
\dot{\gamma} y_2\\
0
\end{array}
\right)
\label{shearflow}
\end{equation}

In the camera method, the particle and the fluid interact only inside the frame of the camera. The camera frame is a disk centered on the particle and has a radius of $10 \ d_p$. The boundary conditions for the fluid on the camera frame boundaries are given by (\ref{shearflow}). A scheme of the model is shown on figure \ref{2Dpartshear}.

The two parameters that drive the system physics are the Reynolds number $Re = \frac{\rho_{ext} \dot{\gamma} d_p^2}{\eta_{ext}}$ and the capillary number $Ca = \frac{\eta_{ext} \dot{\gamma}}{G}$, as shown in \cite{GaoHu}. We fixed the Reynolds number to $Re=0.05$, typical for mimicking low flow regime \cite{GaoHu}. We focused on the role of the capillary number and we made it range from $0.02$ to $0.7$ by adjustments of the fluid shear rate $\dot{\gamma}$.

The particle is known to deform into an ellipse and the deformation can be represented with a dimensionless number built from the lengths of the two axises of the ellipses: $a$ is the largest axis and $b$ is the small axis \cite{GaoHu}:

\begin{equation}
D = \frac{a-b}{a+b}
\label{defD}
\end{equation}

\begin{figure}[h!]
\centering
\begin{minipage}[c]{0.46\linewidth}
\includegraphics[width=6cm]{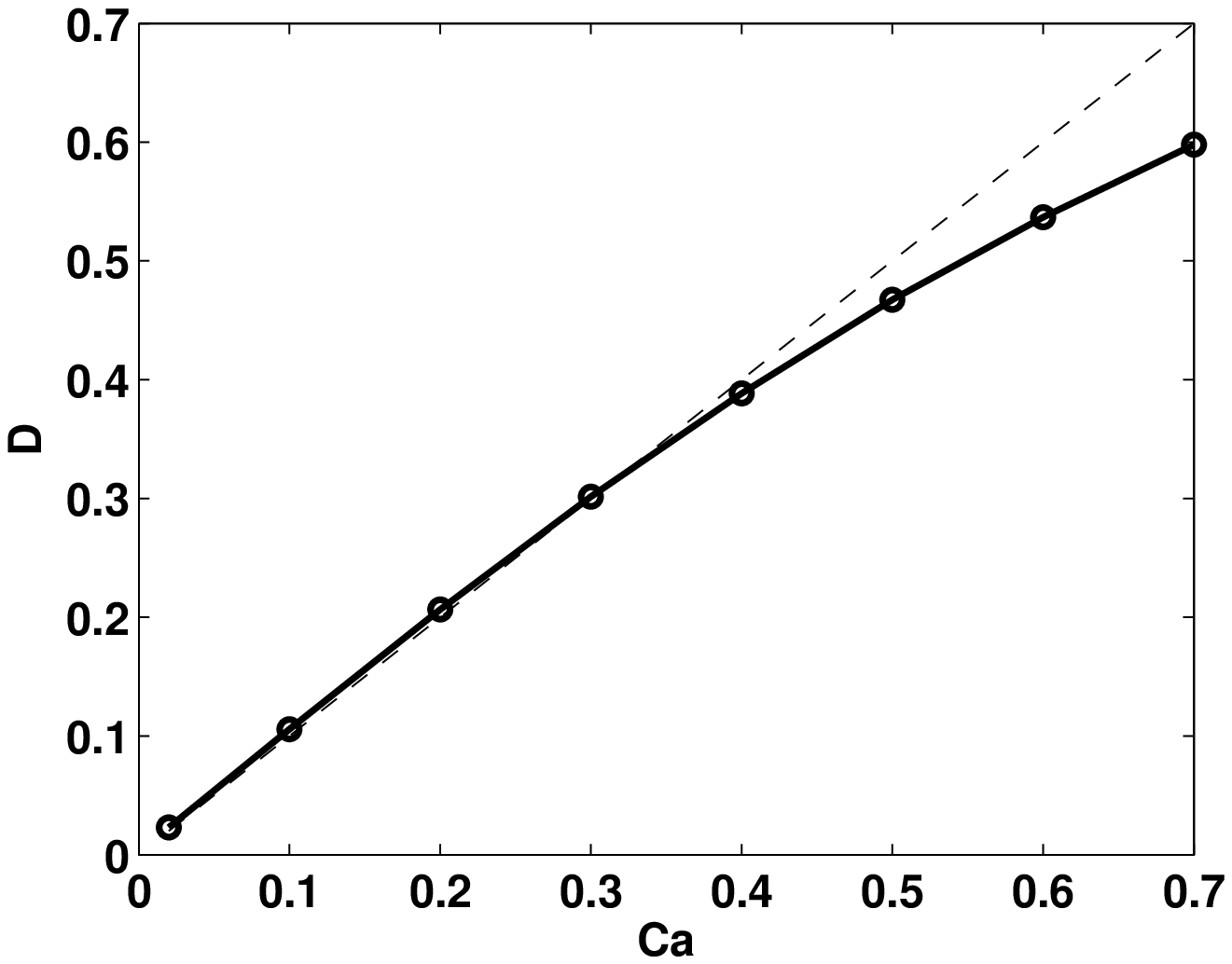}
\end{minipage}
\begin{minipage}[c]{0.46\linewidth}
\begin{tabular}{cc}
\includegraphics[width=3cm]{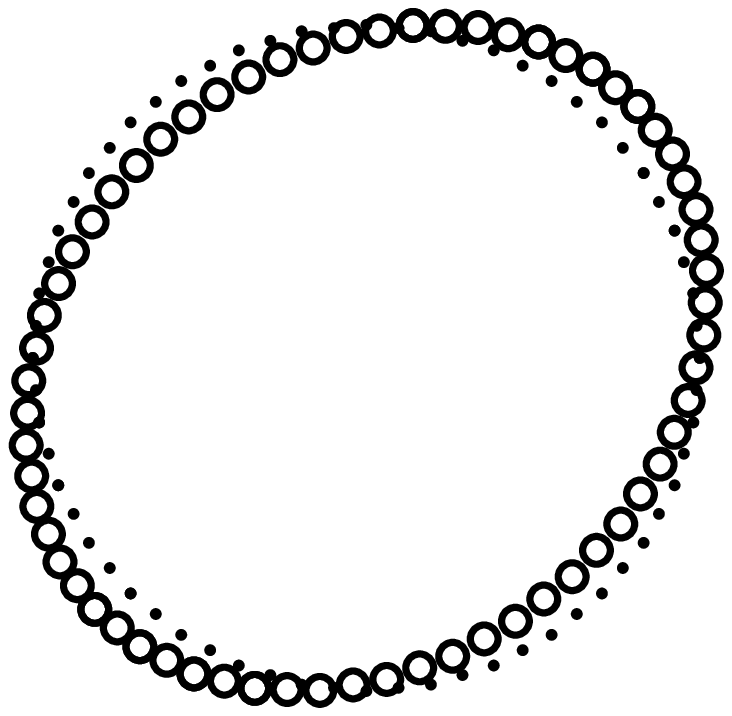} & \includegraphics[width=3cm]{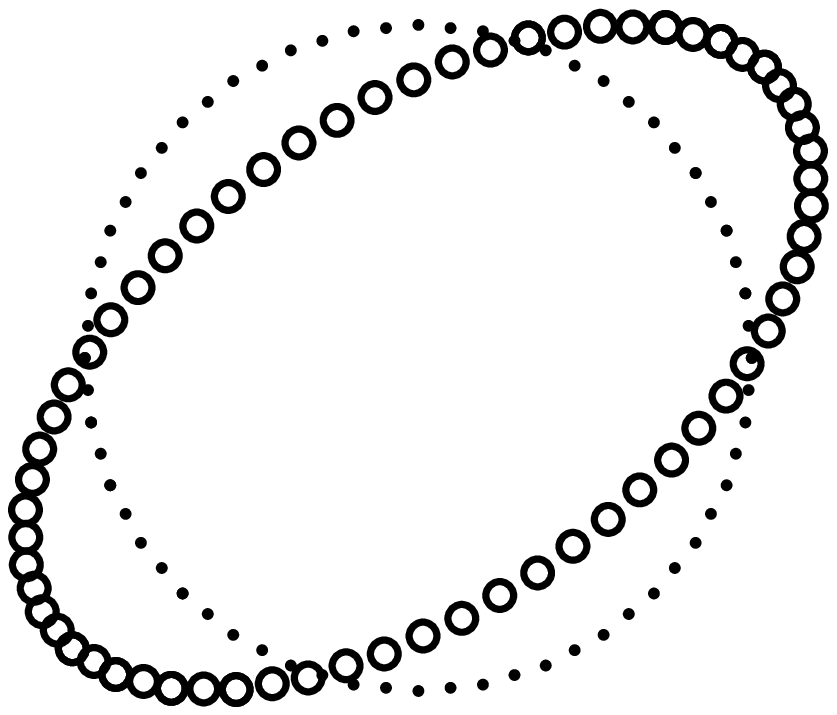}\\
$Ca=0.1$ & $Ca=0.3$\\
\includegraphics[width=3cm]{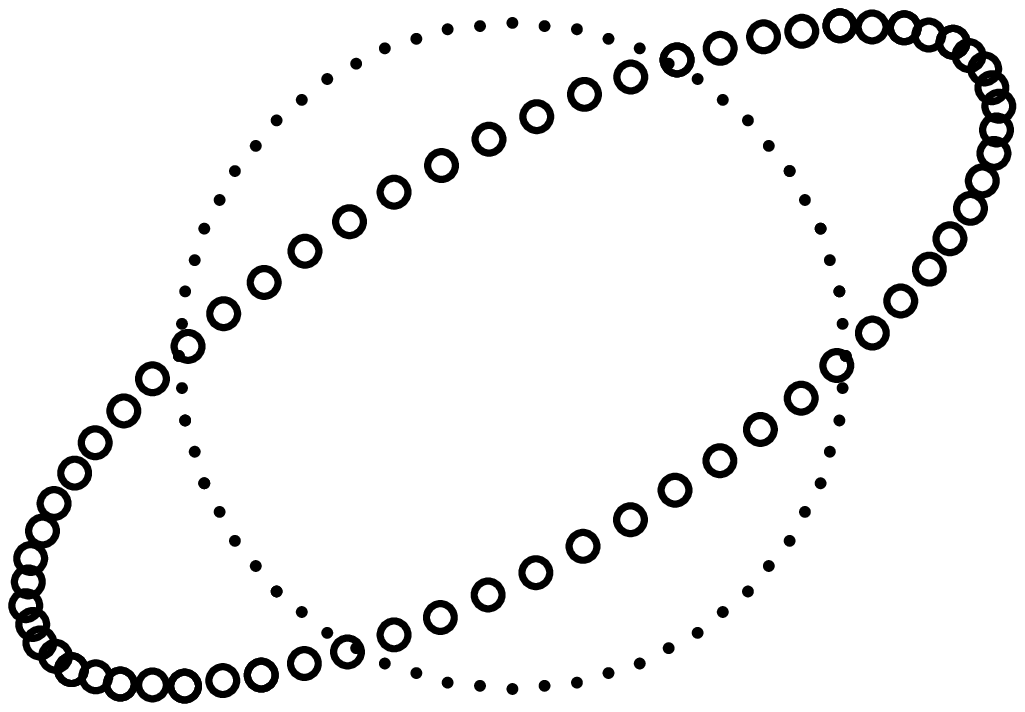} & \includegraphics[width=3cm]{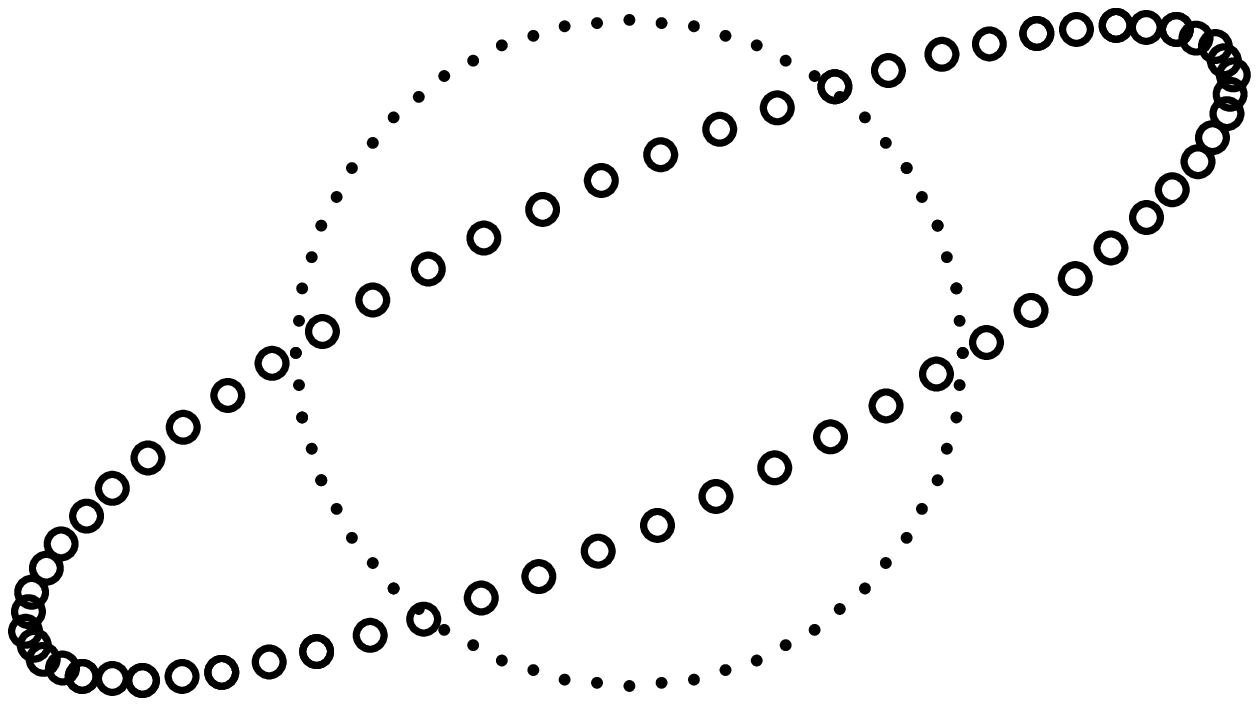}\\
$Ca=0.5$ & $Ca=0.7$
\end{tabular}
\end{minipage}
\caption{Left: particle diameter relative difference $D$ versus its capillary number $Ca$, $Re=0.05$. Right: the circles represents the stationary particle shape and the dots the initial particle shape.}
\label{2DpartshearDvsCa}
\end{figure}

\begin{figure}[h!]
\centering
\begin{psfrags}
\includegraphics[width=6cm]{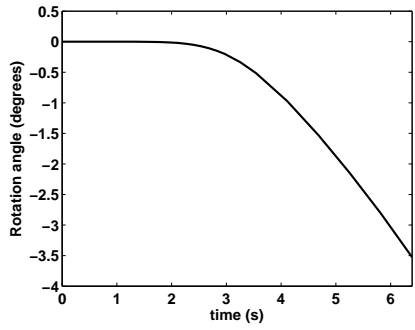}
\end{psfrags}
\caption{Angle (degrees) variation along time for a particle with $Ca=0.4$ and $Re=0.05$. The particle is rotating, with the camera method, the mesh quality remains very good and no remeshing was needed.}
\label{2DpartshearAngle}
\end{figure}

\begin{figure}[h!]
\centering
\begin{psfrags}
\psfrag{X_}[c]{$X$}
\psfrag{Y_}[c]{$Y$}
\psfrag{rot_}[c]{$\dot{\theta}$}
\includegraphics[width=5cm]{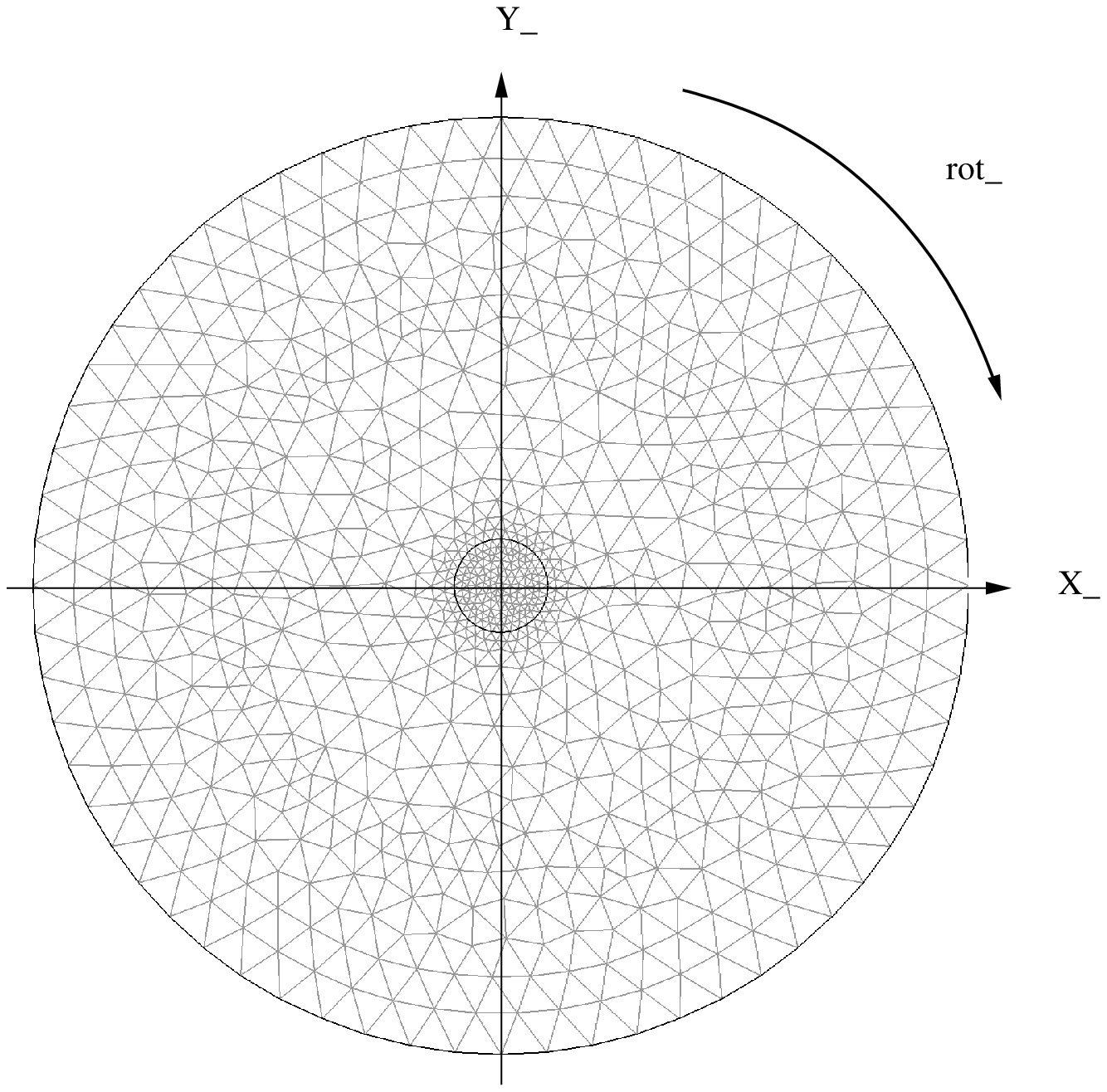}
\includegraphics[width=5cm]{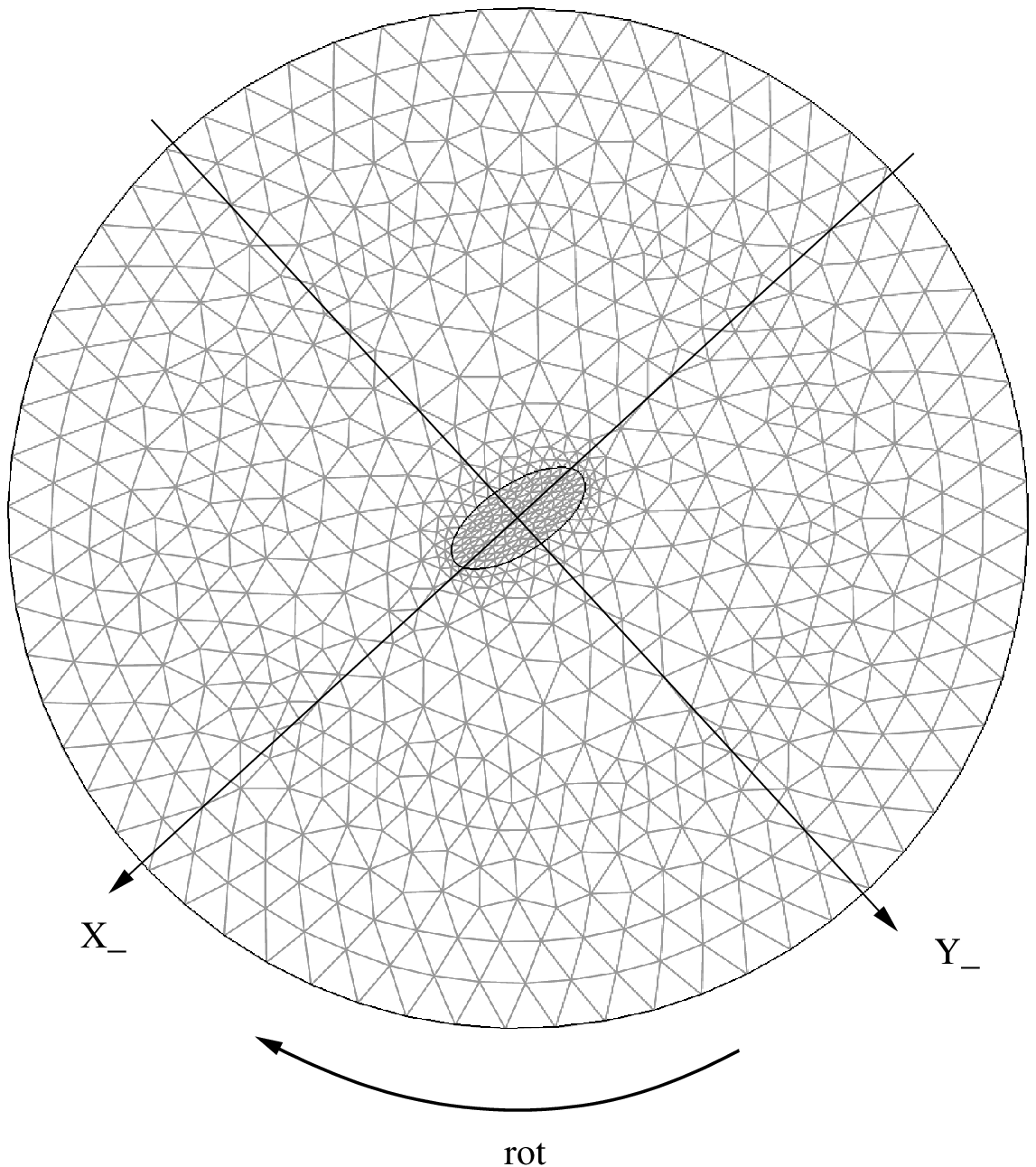}
\end{psfrags}
\caption{Mesh details in the case $Ca=0.4$ and $Re=0.05$. Left: initial ($t=0$). Right: stationary shape ($t=100$). The mesh rotates with the particle.}
\label{2Dmeshrot}
\end{figure}

Figure \ref{2DpartshearDvsCa} shows the results obtained with the camera method. The Reynolds number is $Re=0.05$ in all simulations. The variations of $D$ versus the capillary number $Ca$ are plotted on the left. As expected, the number $D$ follows a linear regime $D=Ca$ for values of $Ca$ lower than $0.3$-$0.4$ and starts to shift downwards the curve $D=Ca$ when $Ca$ increases. 

Once in stationary regime, the particle rotates around its gravity center at a constant velocity speed of $\dot{\theta}=-1.42 \ degrees.s^{-1}$ which is independent of $Ca$, the camera method makes the mesh rotate with the particle and avoid  triangle elements to distort too much. Angle variation is plotted on figure \ref{2DpartshearAngle} in the case $Ca=0.4$. The minimal quality of the mesh elements during the computation remains in that case always larger than $0.55$. Figure \ref{2Dmeshrot} gives an example on how the camera mesh rotates and deforms and shows clearly that remeshing is not necessary. 

\subsubsection{A particle in a bifurcation}

\begin{figure}[h!]
\centering
\begin{psfrags}
\psfrag{initial_}[l]{Initial position}
\psfrag{time_}[l]{Position at time $0.04 \ s$}
\includegraphics[width=9cm]{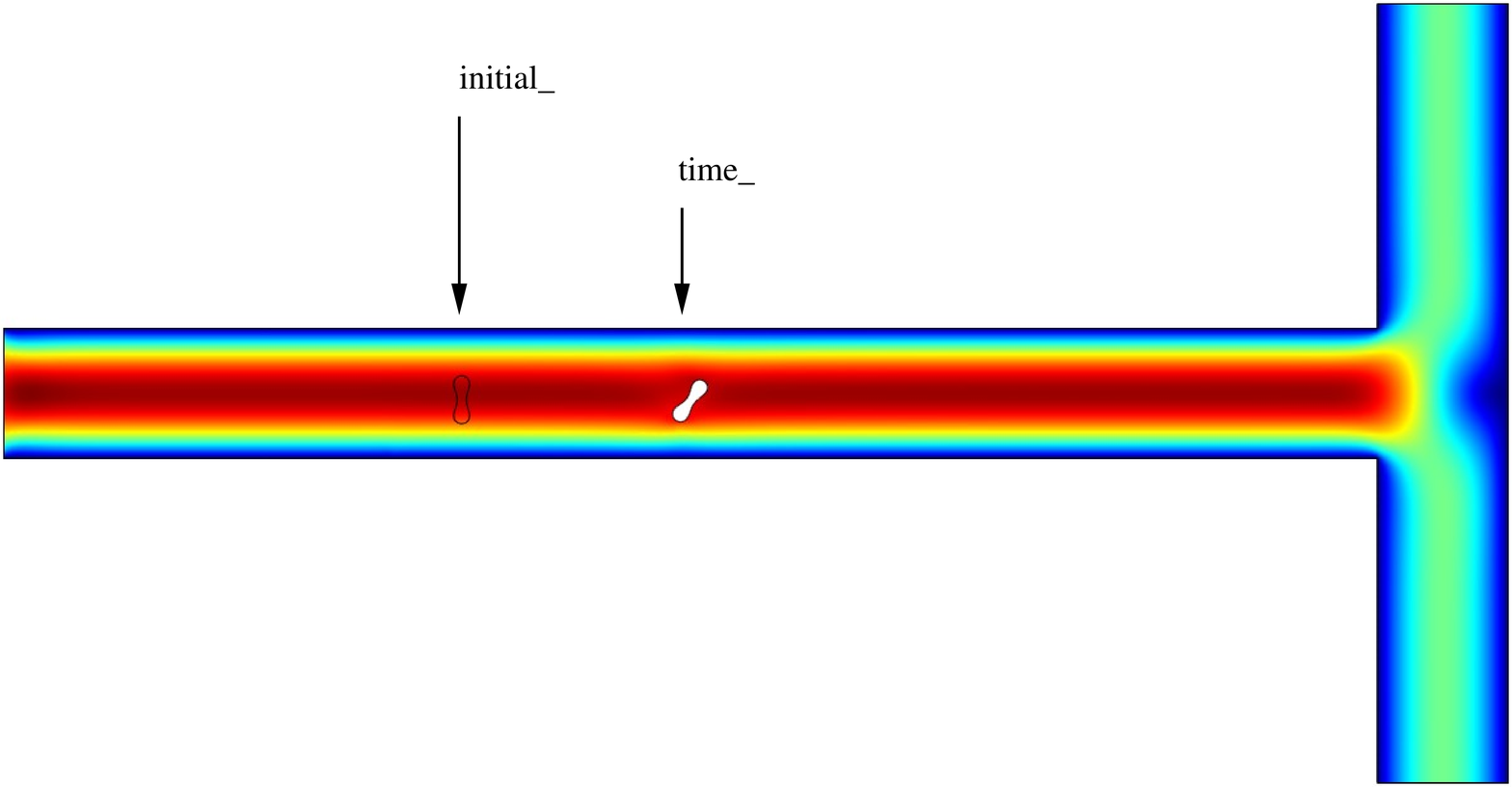}
\includegraphics[width=5cm]{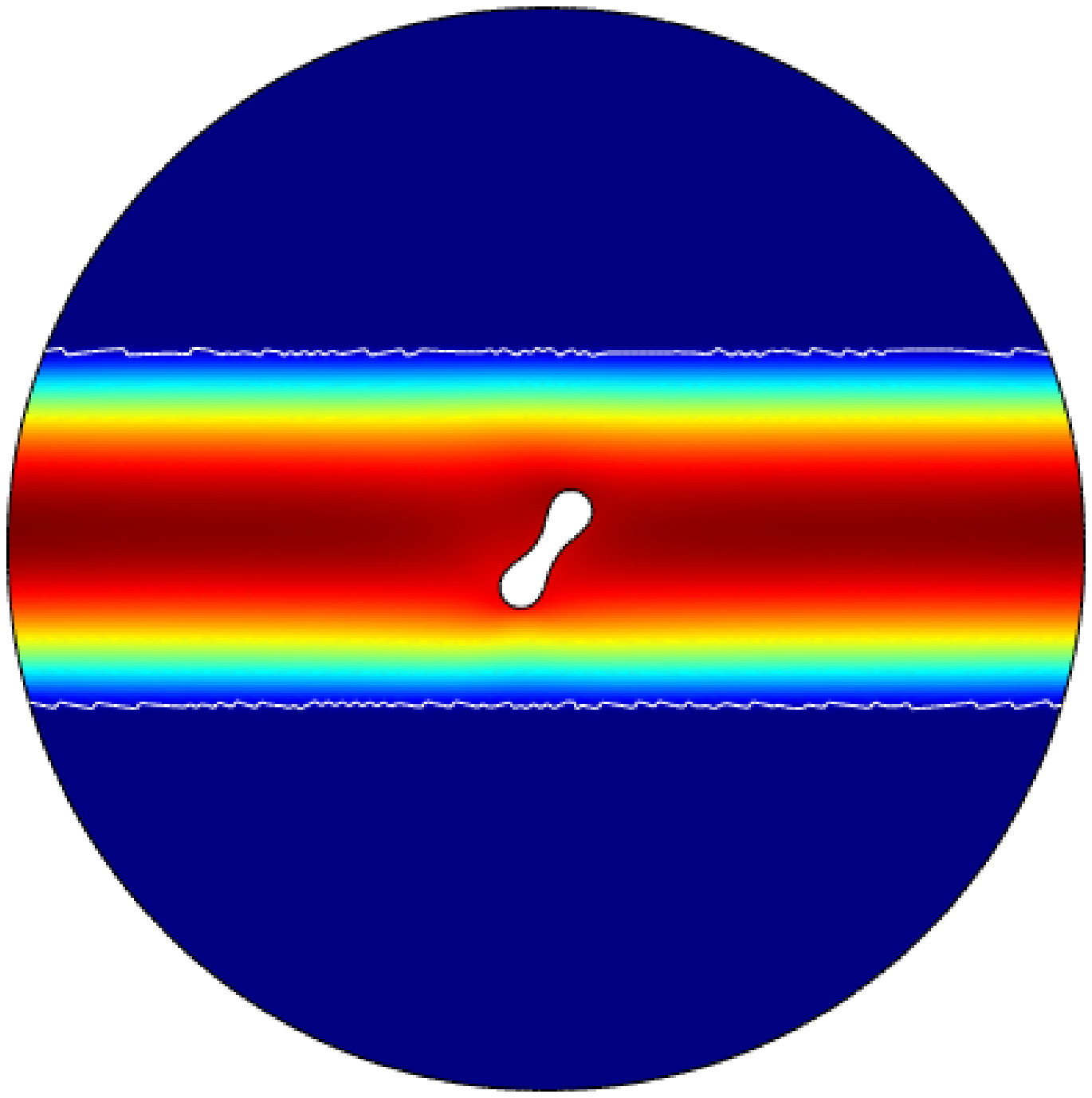}
\end{psfrags}
\caption{The fluid domain is a bifurcation, the channels diameter is $20$ microns. The fluid enters the network from the inlet on the left and gets out from the two outlets on the right (up and down). The particle is initially positioned near the inlet. The color represents the amplitude of the fluid velocity (increasing from blue ($0 \ mm.s^{-1}$) to red ($1 \ mm.s^{-1}$)). The camera frame on the right is a disk of diameter $80 \ \mu m$ centered on the particle, the image represents the particle at time $0.04 \ s$.}
\label{bif2d}
\end{figure}

In this section, we detail the use of the camera method for the case of a particle going through a fluid network shaped as a bifurcation, the network geometry is shown on figure \ref{bif2d} (left). In this example, the wall of the network intersects the camera frame all along the computation, as shown on figure \ref{bif2d} (right). The topology of the network part inside the camera frame changes and the solution that consists in sticking the camera boundary to the network boundary is not any more possible. Thus we use penalization to force fluid velocity to be zero in the camera frame part outside of the network. The diameters of the network channel is constant everywhere and equal to $20 \ \mu m$.

The particle shape and size are those of a section of a red blood cell, its material is elastic (Young's modulus $ 68 \ Pa$ and Poisson's ratio $0.4995$) and undergoes large deformations, its density is that of water. The fluid is assumed to be water at low flow regime (Stokes equations). We assume that two same parabolic velocity profiles are imposed at up and down outlets of the bifurcation. An open boundary condition (zero constraint) is applied at the inlet. The maximal velocity reached on the horizontal channel center line is $0.001 \ m.s^{-1}$.
The particle is initially positioned next to the inlet, one micron below the center line of the flow as shown on figure \ref{bif2d}, and it is carried away in the network by the fluid.

We compared the results of two numerical methods to compute the particle displacements in the bifurcation. In all our simulations, the mesh is made of triangular elements and the mesh size represents the maximal length of triangles edges. The first method is a standard ALE method and the computation is stopped when the mesh quality becomes too bad. The second method is the camera method. We confronted results and computation times.

\begin{figure}[h!]
\centering
\includegraphics[width=15cm,trim=4cm 3cm 2cm 2cm, clip]{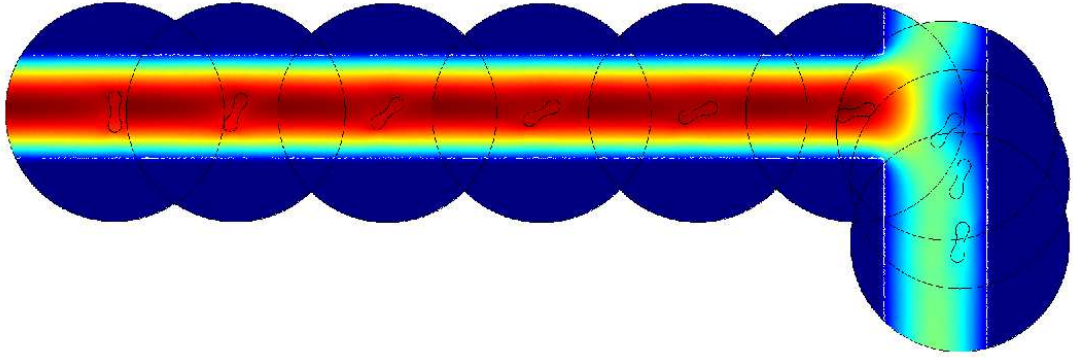}
\caption{Particle trajectory computed with the camera method (camera diameter $60 \ \mu m$). The particle and the camera (circle centered on the particle) move from left to right and are plotted each $0.03 \ s$. The color represents the amplitude of the fluid or structure velocity (increasing from blue ($0 \ mm.s^{-1}$) to red ($1 \ mm.s^{-1}$)). The computation stops a bit before the camera borders cross the bottom outlet.}
\label{camfull}
\end{figure}

\begin{figure}[h!]
\centering
\includegraphics[width=6cm]{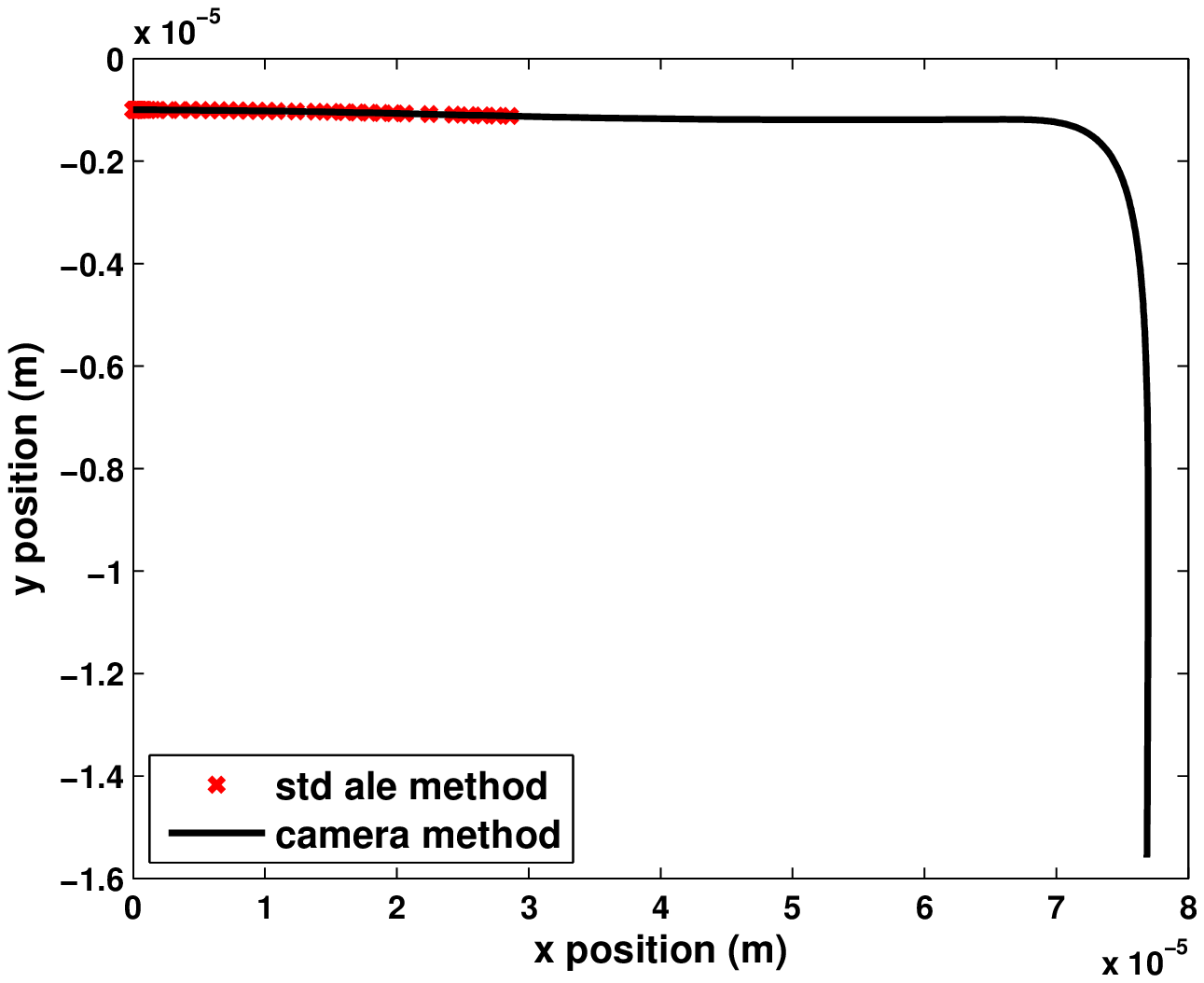}
\includegraphics[width=6cm]{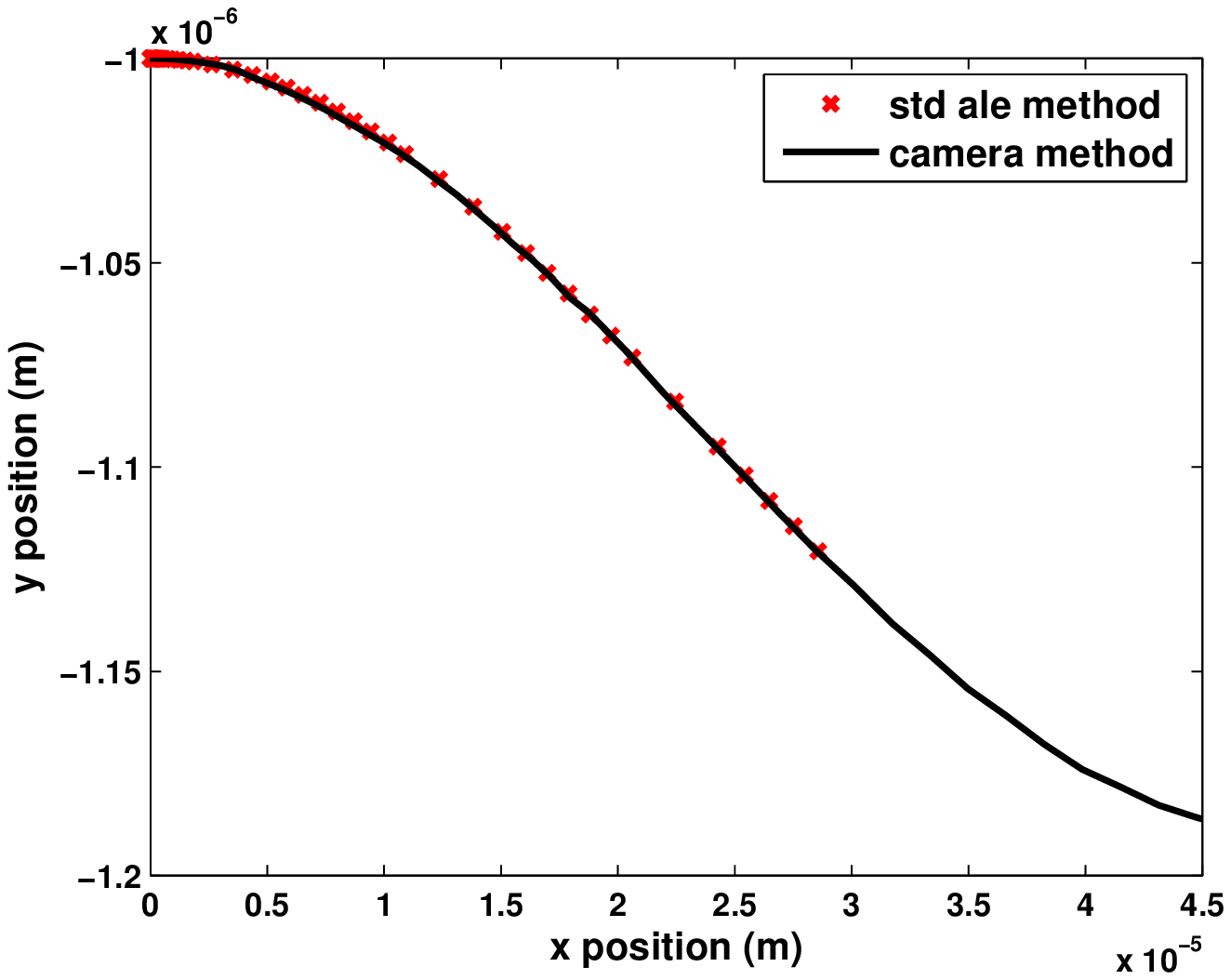}\\
\includegraphics[width=6cm]{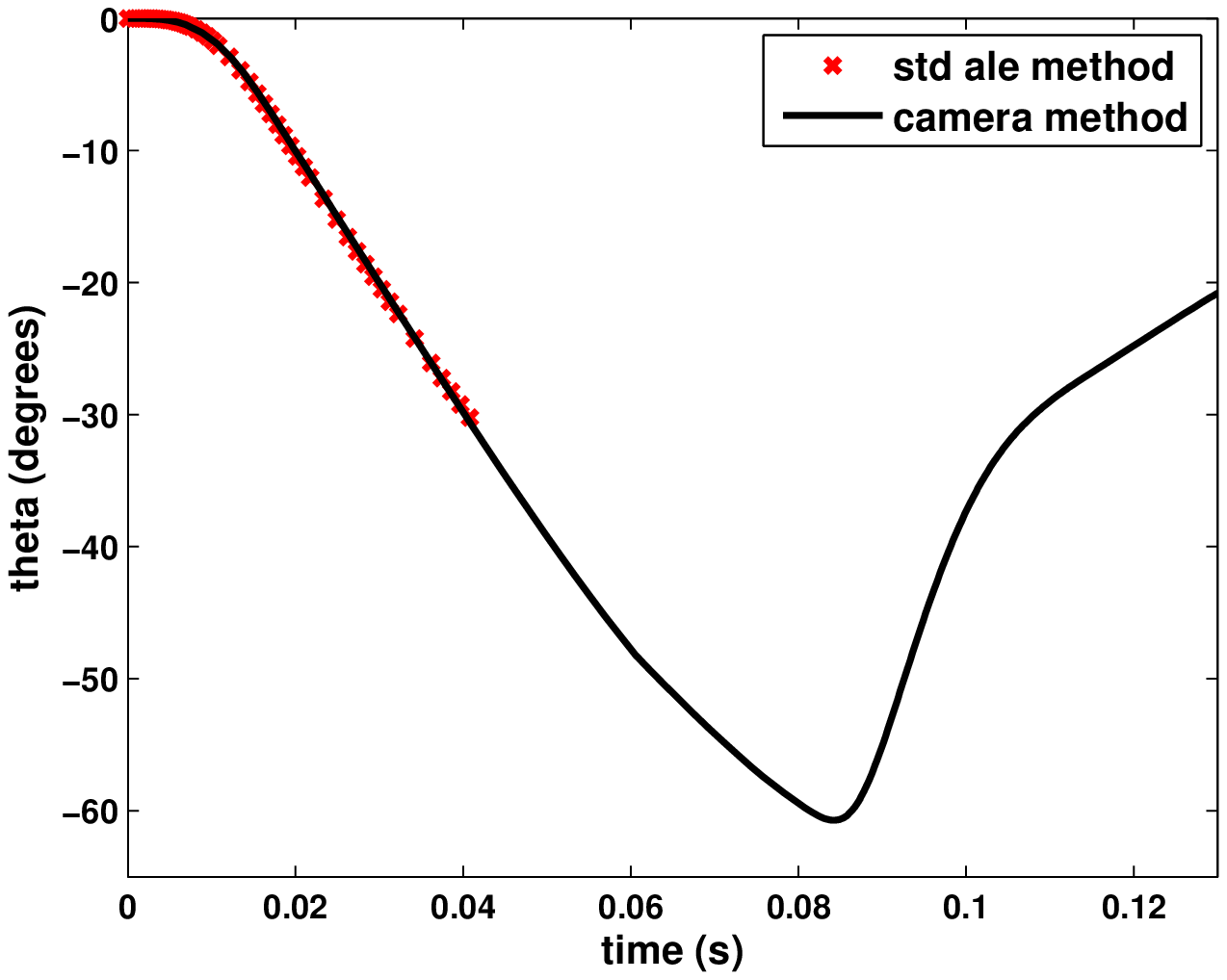}
\includegraphics[width=6cm]{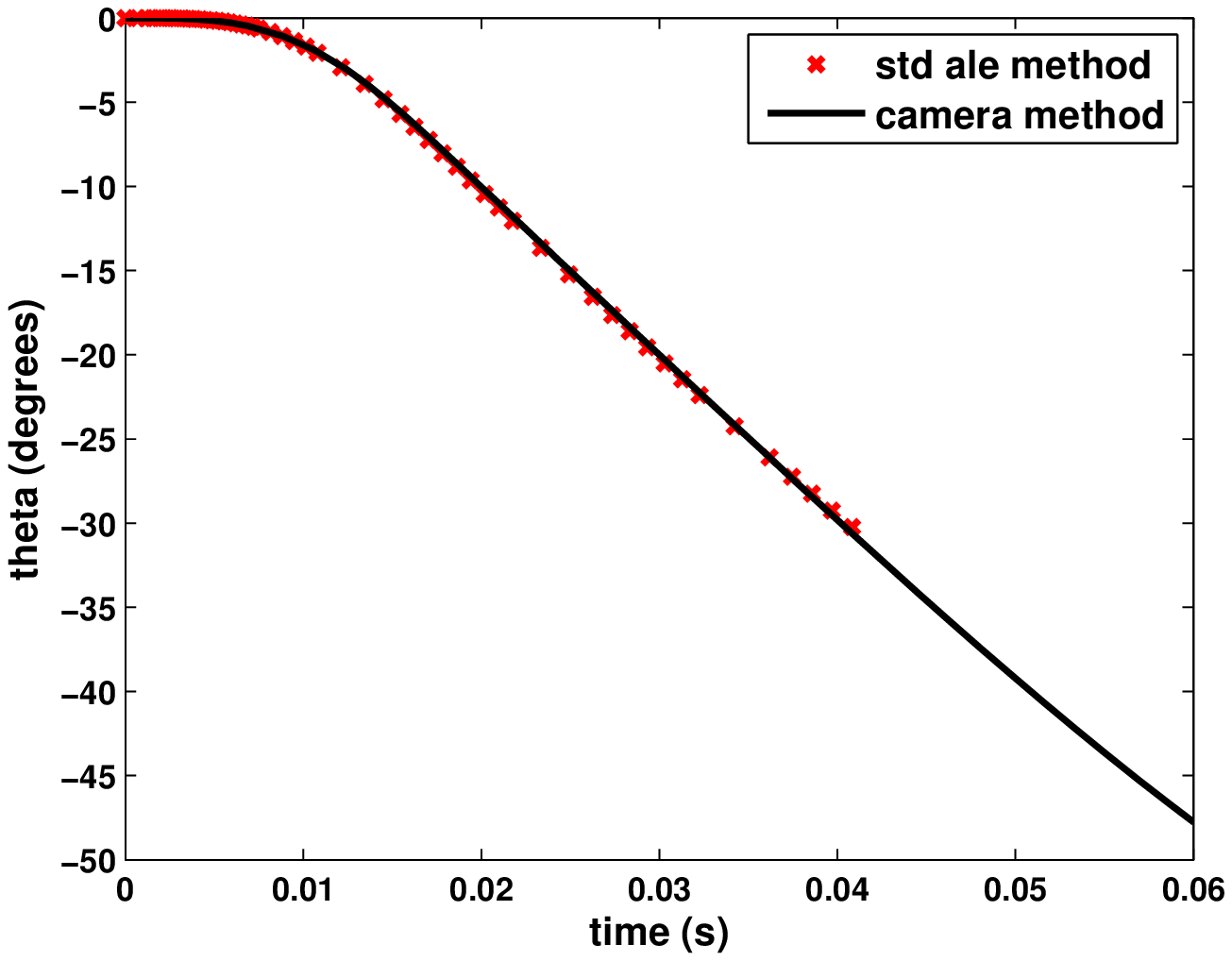}
\caption{Up: particle trajectory (meters), the right plot is a zoom. Down: particle rotation (degrees), the right plot is a zoom. The black lines correspond to the camera method and the red crosses correspond to the standard ALE method. Note that the standard ALE method was not able to simulate the particle after the time $0.04 \ s$.}
\label{compfig}
\end{figure}

With the standard ALE method the simulation is not able to reach the time when the particle enters the bifurcation. On the contrary, the camera method is able to simulate the motion of the particle all the way through the bifurcation, from the inlet to one of the outlets, with very few loss in mesh quality, see figure \ref{camfull}. We compared the particle trajectories and rotations of the camera method with the standard ALE method using the same mesh size $1 \ \mu m$ and a camera frame diameter of $60 \ \mu m$. The particle trajectory and rotation computed with the camera method are shown on figure \ref{camfull} and they are compared with the standard ALE method on figure \ref{compfig}.

\begin{figure}[h!]
\centering
\includegraphics[width=6cm]{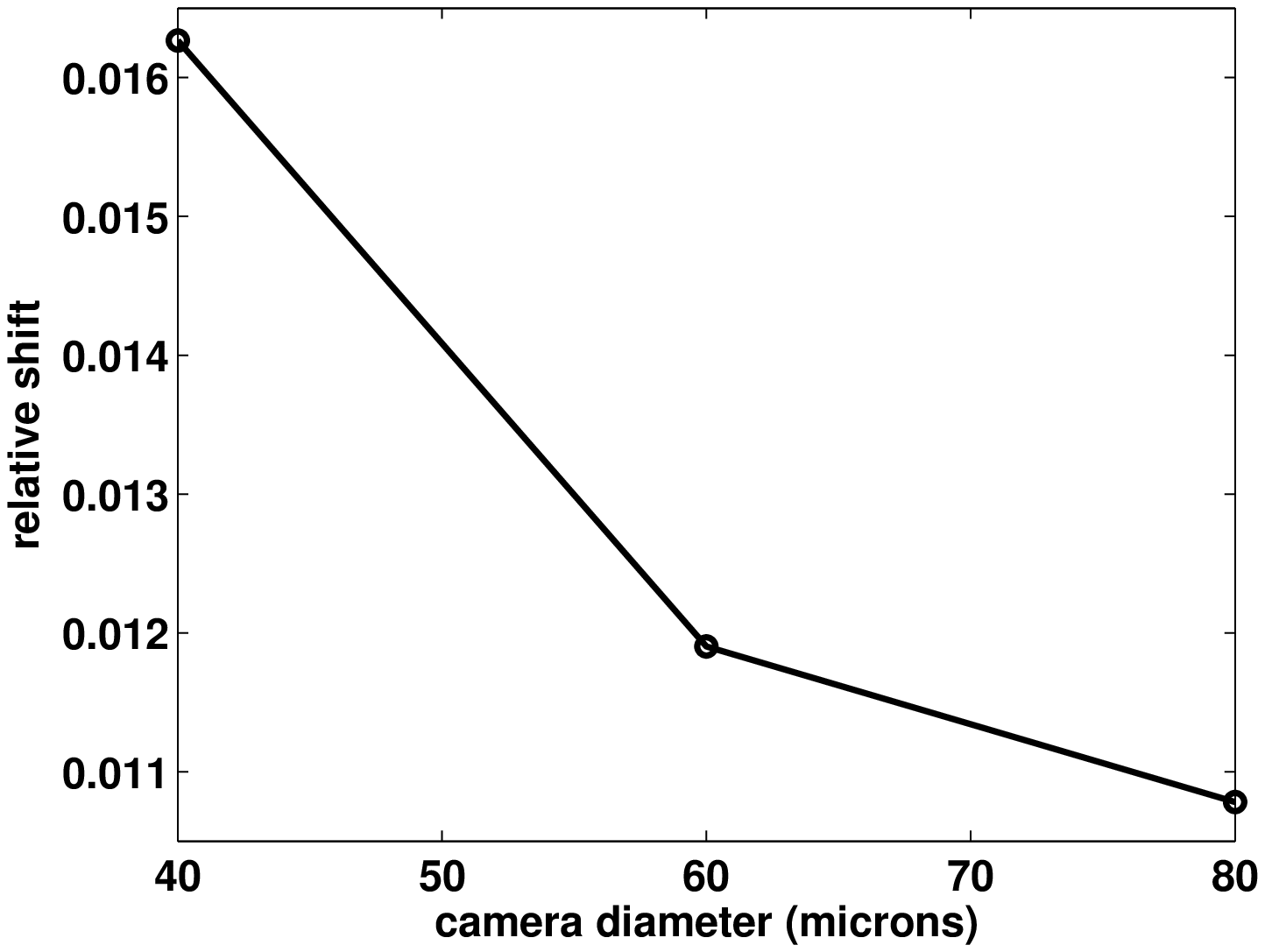}
\includegraphics[width=6cm]{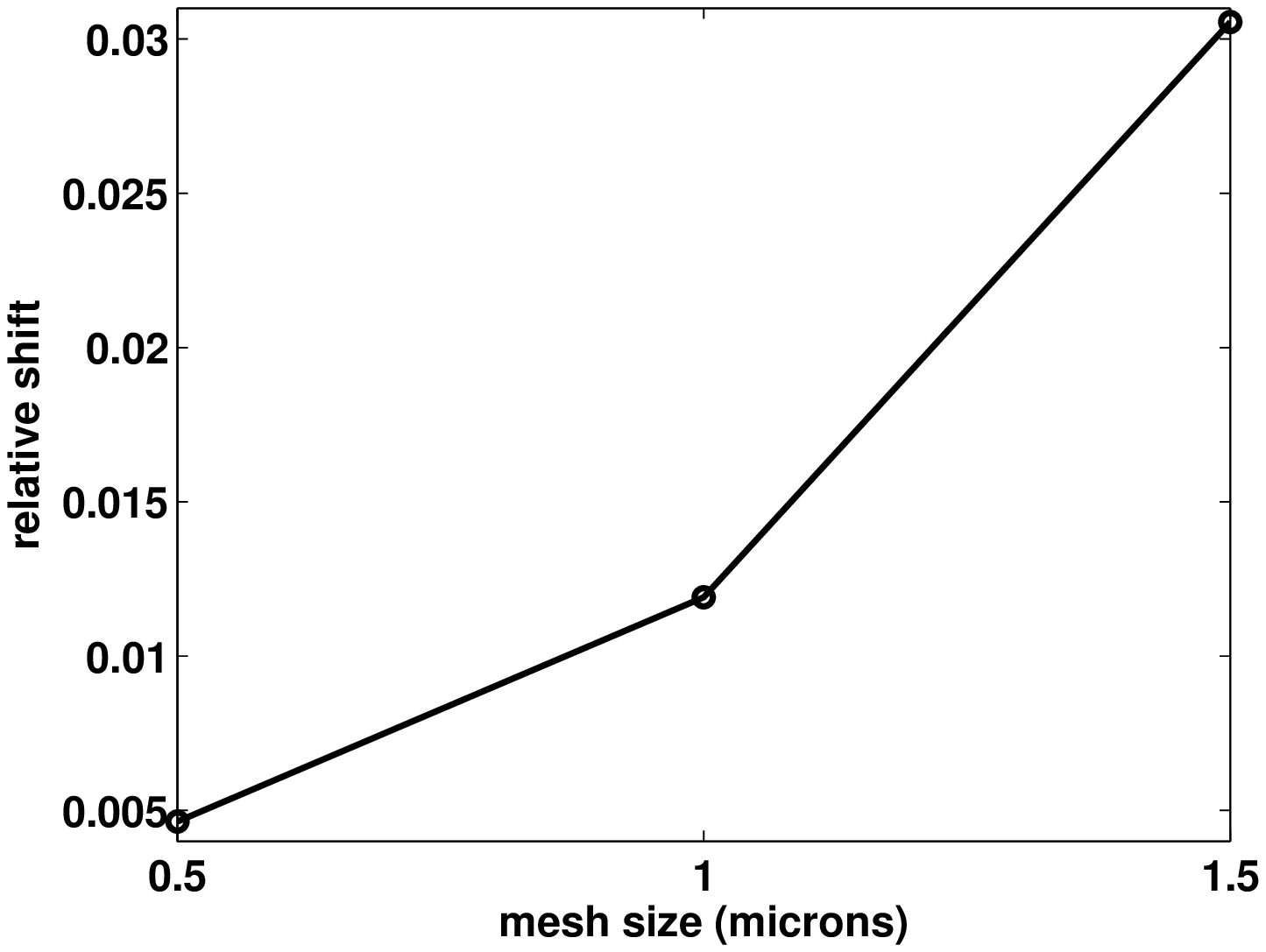}
\caption{Left: relative shift of particle position $\mathcal{S}$ for different sizes of the camera frame (mesh size $1 \ \mu m$). Right: relative shift of particle position $\mathcal{S}$ for different mesh sizes (camera frame diameter is $60 \ \mu m$).}
\label{framesizeandmesh}
\end{figure}

We compared standard ALE method and camera method by measuring the difference in the particle gravity center displacement $(dx, dy)$ and particle rotation $\theta$ at time $t=0.04 \ s$, the last time computed by the standard ALE method before it stops due to low elements quality. The relative shift $\mathcal{S}$ on particle position is computed relatively to the results given by the standard ALE method with a fine mesh (mesh size $0.5 \ \mu m$). The relative shift includes one term for each coordinate of particle position and one term for particle rotation: 

$$
\mathcal{S}= \frac13 \sqrt{\frac{(dx_{cam}-dx_{ale})^2}{dx_{ale}^2} + \frac{(dy_{cam}-dy_{ale})^2}{dy_{ale}^2} 
+ \frac{(\theta_{cam}-\theta_{ale})^2}{\theta_{ale}^2}}
$$

Since the fluid velocities applied on the camera frame boundaries are those computed in the absence of particle, their quality as approximations is all the more better than they are taken far from the particle, so the camera frame size plays a role on that quality. 

The penalization term used to define the walls of the channel in equation \ref{CAMfluid} involves the characteristics function $\chi$ of the channel walls, i.e. $\chi(x) = 1$ if the point $x$ of the camera at time $t$ is in the channel wall and $\chi(x) = 0$ if $x$ is in the channel. To ensure convergence of the computation, it is necessary to smooth the function $\chi$ and for the smooth to be meaningful, it has to span on at least two elements. So the approximation of the function $\chi$ and consequently approximation of the wall position is all the more accurate than the mesh size is small.

\begin{table}[h!]
\centering
$$
\begin{array}{|l|c l|c|c|c|}
\hline
\text{Method} & \begin{array}{c}\text{Elements}\\ \text{number} \end{array} & \text{(mesh size)} &  \begin{array}{c} \text{Frame}\\ \text{diameter} \end{array} & \begin{array}{c} \text{Simulation}\\ \text{time} \end{array} &  \begin{array}{c} \text{Relative}\\ \text{shift $\mathcal{S}$} \end{array}\\
\hline
\text{\bf{std. ALE}} &&&&&\\
\text{\vspace{20 mm} fine mesh} & 67 826 & (0.5 \ \mu m) & N/A & 5301.7 \ s & 0 \ \% \text{ (ref)}\\
\hline
\text{\bf{camera}} &&&&&\\
\text{\vspace{20 mm} normal} & 10160 & (1 \ \mu m) & 60 \ \mu m & 491.2 \ s & 1.19 \ \%\\ 
\text{\vspace{20 mm} small frame} & 5222 & (1 \ \mu m) & 40 \ \mu m & 244.6 \ s & 1.63 \ \%\\ 
\text{\vspace{20 mm} large frame} & 17374 & (1 \ \mu m) & 80 \ \mu m  & 967.5 \ s & 1.08 \ \%\\
\text{\vspace{20 mm} coarse mesh} & 3662 & (1.5 \ \mu m) & 60 \ \mu m & 232.8 \ s & 3.06 \ \%\\
\text{\vspace{20 mm} fine mesh} & 35998 & (0.5 \ \mu m) & 60 \ \mu m & 3135.0 \ s & 0.46 \ \%\\
\text{\vspace{20 mm} fine mesh $\&$ small frame}& 16 914 & (0.5 \ \mu m) & 40 \ \mu m & 1088.3 \ s & 0.55 \ \%\\
\hline
\end{array}
$$
\caption{Simulations data and computation times.}
\label{computdata}
\end{table}

Thus, we tested the influence on the relative shift $\mathcal{S}$ of the camera frame diameter and of the mesh size. The results are plotted on figure \ref{framesizeandmesh} and computations properties and times are reported on table \ref{computdata}. As expected, an increase of the camera frame size or of the number of mesh elements reduces the shift. The curves show that a large part of the shift comes from the smooth of the $\chi$ function involved in the penalization term and a decrease of mesh size is more effective in term of shift reduction than an increase of the camera frame size. Consequently, a small camera frame is very efficient, computation time is small and quality remains good. Thus, reducing the camera frame is a good strategy as long as the camera remains large enough to capture most of the effects of the particle on the fluid.  To illustrate the previous points, a computation was made with a small camera frame (diameter $40 \ \mu m$) and a fine mesh (mesh size $0.5 \ \mu m$), the relative shift found was $0.55 \ \%$ and the computation time was five times smaller than the standard ALE method.

\subsection{3D}

In this section, we describe how to implement the camera method for three-dimensional particle and fluid. We use the coordinates $x=(x_1, x_2, x_3)$ for the reference frame and $y=(y_1, y_2, y_3)$ for the deformed frame.

\label{3D}

\subsubsection{Specificity of the camera method in 3D}

In a three dimensional space, a rotation is defined with three angles. Different ways to define the rotations are possible and we chose to use Euler angles \cite{Salomon}. We will write $\theta(t) = \left( \theta_1(t), \theta_2(t), \theta_3(t) \right)$. With this definition for rotations, the matrix $R_{\bf \theta(t)}$ is the product of three 3D rotation matrices:

$$
R_{\bf \theta(t)} = R_{\theta_3(t)} R_{\theta_2(t)} R_{\theta_1(t)}
$$

$R_{\theta_i}$ represents the rotation around the axis $i$ whose angle is $\theta_i$. In 3D, it is not possible to calculate an analytical expression for the three angles of rotations. Thus, the three angles are computed numerically by solving the three non linear equations arising from the rotation constraint on the elementary displacement $\int_{S_0} x \wedge d(x,t) dx = 0$, which can be rewritten with the full displacement $u$: $\int_{S_0} x \wedge R_{-\theta(t)}\left( x + u(x,t) - \tau(t) \right) dx = 0$. These equations are coupled to the fluid-structure interaction equations in the camera and solved together.

\begin{figure}[h!]
\centering
\begin{psfrags}
\psfrag{x_}{$x_1$}
\psfrag{y_}{$x_2$}
\psfrag{z_}{$x_3$}
\psfrag{camera frame_}[l]{camera frame}
\psfrag{vesicle_}{particle}
\includegraphics[width=6cm]{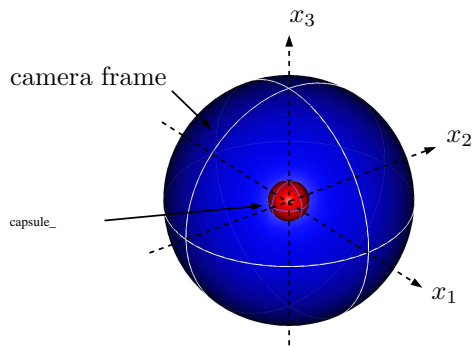}
\end{psfrags}
\caption{Example of a typical camera frame in 3D space. The camera is a sphere centered on the particle and filled with fluid. The fluid boundary conditions on the boundary of the camera frame are given either by analytical data or by a priori numerical simulations of the fluid in the whole network without the particle. The camera moves and rotates with the particle.}
\label{3Dcamframe}
\end{figure}

Figure \ref{3Dcamframe} shows a typical camera frame, a sphere centered on the particle. As for the bi-dimensional case, the shape of the camera frame can be any shape enclosing the particle and wide enough to be able to neglect the effect of the particle on the fluid outside the camera frame.

\subsubsection{3D validation: a vesicle in shear stress}

In this section, we consider a 3D spherical vesicle laying on the center line of a shear flow. More precisely, we assume that the fluid velocity without the vesicle (or far from the vesicle) is a pure shear flow. Thus, if $\dot{\gamma}$ is the shear rate of the fluid, its analytical expression is:

\begin{equation}
v(y)=\left(
\begin{array}{c}
0\\
\dot{\gamma} y_3\\
0
\end{array}
\right)
\end{equation}

A 3D spherical vesicle on the center line of a shear flow deforms into an ellipsoid in the direction defined by the shear flow velocities \cite{Gong, Egg, Lac}. The properties of the vesicle deformation depend on the capillary number $Ca$, which is computed from the shear rate of the fluid, the viscosity of the outer fluid $\eta_{out}$ and the shear modulus of the membrane $G$:
 
$$
Ca=\frac{\eta_{out} \dot{\gamma}}{G}
$$

\begin{figure}[h!]
\centering
\begin{psfrags}
\psfrag{a_}{$a$}
\psfrag{b_}{$b$}
\includegraphics[width=4cm]{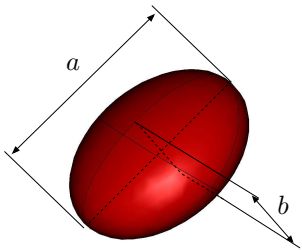}
\end{psfrags}
\caption{Diameters $a$ and $b$ used to compute the dimensionless number $D_{yz} = \frac{a-b}{a+b}$  for the $3D$ vesicle.}
\label{Dxy3D}
\end{figure}

Then, the deformation of the vesicle can be represented with a dimensionless number $D_{yz}$ built with the largest diameter $a$ and the smallest diameter $b$ of the ellipsoid in the plane defined by the shear flow velocities, here $(yz)$, see figure \ref{Dxy3D}:  

$$
D_{yz} = \frac{a - b}{a + b}
$$

In linear regime, i.e. for small capillary number, the dependence of $D_{yz}$ with the capillary is known analytically, and $D_{yz}$ varies linearly with $Ca$ \cite{pozrikidis}:

$$
D_{yz} = \frac54 \frac{2+\nu}{1+\nu} \ Ca
$$

In non linear regime, the dependence of $D_{yz}$ with the capillary number has been studied numerically in the literature using thin membrane models, see for example \cite{Barthes1,Barthes2, Lac}. Thus, we simulated the deformation of a spherical vesicle in a shear flow with the 3D camera method and we compared the $D_{xy}$'s  computed with the camera method with the $D_{xy}$'s computed in the literature \cite{Gong, Egg, Lac}. We assume that the membrane of the vesicle is made of a thick Neo-Hookean hyperelastic material and that the thickness of the membrane is $2.5 \%$ the vesicle diameter. The camera frame is a sphere centered on the vesicle, the diameter of the sphere is five times the diameter of the vesicle, see figure \ref{3Dcamframe}. As in the preceding section, we assume that the vesicle does not affect the fluid outside of the camera frame. The analytical expression of the velocity gives fluid boundary conditions on the camera boundaries.

\begin{figure}[h!]
\centering
\includegraphics[width=10cm]{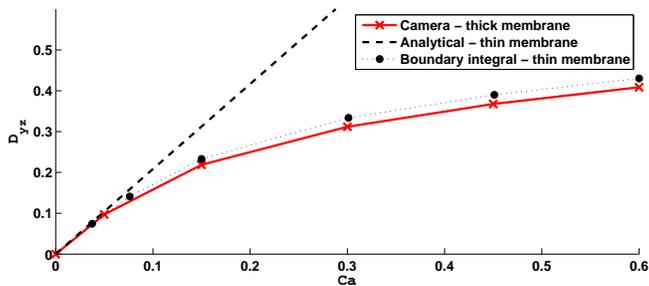}
\caption{$D_{yz}$ versus capillary number. The dashed line represents the analytical expression of $D_{yz}$ for small deformations, the dotted line represents the boundary integral simulations with an infinitely thin membrane from \cite{Barthes1}, the continuous line represents the numerical simulations with the camera method using a thick membrane model.}
\label{Dxyvscapnum}
\end{figure}

Results are plotted on figure \ref{Dxyvscapnum}. The red continuous curve represents the aspect ratios computed with the camera method and the thick membrane; the black dotted curve represents the results computed by \cite{Barthes2} with a thin membrane model; the dashed line represents the linear case. As expected, for low capillary numbers ($Ca < 0.05$) the vesicle behaves linearly. For higher capillary numbers, the aspect ratio is no more linear. Our results are very close to that of \cite{Barthes2} except for a slight downward shift that should be expected because of the differences between the membrane models. Indeed, bending forces are not accounted for in that particular thin membrane model, but they are accounted for in thick membrane models such as ours. Bending forces withstand to the membrane displacements and models neglecting them slightly overestimate $D_{yz}$. Nevertheless, bending forces are often neglected because most of the time their effects remain small, and this is confirmed by our results for this case.

\section{Discussion}
\label{discussion}

The camera method enables to compute the motion and deformation of a particle in a fluid domain of any size. The simulations are performed on a neighbourhood of the particle only, consequently the size of the numerical problem is drastically reduced. Moreover, the mesh moves and rotates with the particle, which avoids most of the remeshings. This spares computation time and avoids a potential loss of precision relative to the successive projections of the numerical solutions on new meshes. Finally, the camera method can be easily implemented with any finite elements library and any non-linear solver, such as Comsol Multiphysics. The camera method is particularly adapted to study the details of the behaviour of a single particle moving in a network.

Fluid-structure interaction problems are inherently non linear, and the camera method adds non linearity to the system by adding in the equations the instantaneous rotation angles and translation vector. However, one can avoid this new non linearity in the numerical scheme by using explicit rotation angles and translation vector. Using the rotation angles and translation vector of the preceding time step induces an approximation and a mesh distortion that depends on the time step chosen. The time step can be easily tuned however, for example using linear time estimation of particle translation and rotations:
$$
\begin{array}{l}
\tau(t_{n+1}) = \tau(t_n) + \left( t_{n+1}- t_n \right) \frac{d \tau}{dt}(t_n)\\
\theta(t_{n+1}) = \theta(t_n) + \left( t_{n+1}- t_n \right) \frac{d \theta}{dt}(t_n)\\
\end{array}
$$
If the estimated translations and rotations for the time we want to compute ($t_{n+1}$) remain "close" to that of the preceding time ($t_n$), then the mesh distortion induced by the use of explicit formulation remains reasonable. Consequently, the time step chosen will have to depend on the particle mean velocity and particle mean rotation velocity at the preceding time ($t_n$).

Working with only a subpart of the fluid domain makes necessary to determine fluid boundary conditions on the boundaries of the camera frame. Ideally, we would like to apply exact boundary conditions. Uniqueness  theorems for the solution of Stokes equations or for Navier-Stokes equations (at least at low Reynolds numbers) \cite{Simon} would then ensure that the fluid structure interaction problem is not altered by the domain restriction. Unfortunately we are not able to determine exact boundary conditions since it would require to solve the whole problem, which is exactly what we intend to avoid. Consequently, we need to find approximate boundary conditions. Fluid properties near the particle are highly perturbed and too complex to be easily predicted, but far from the particle, the particle is seen by the fluid as a simple extra pressure drop. In these regions only, we can hope to approximate correctly the behaviour of the fluid. The first consequence for the camera frame is that it has to be wide enough so that its boundaries do not cross the perturbed fluid. Next, two types of boundary conditions for the fluid on the camera frame are possible: either fluid constraint conditions (Neumann, $\sigma_f(v,p).n = g_0$) or velocity conditions (Dirichlet, $v=v_0$). Fluid constraints are always strongly dependent on the particle behaviour and position since the particle affects the pressure distributions globally in the network. On the contrary, velocities can be only weakly dependent on the particle behaviour thanks to flow conservation ($div(v)=0$) and in this case, its dependence is vanishing when going away from the particle \cite{Galdi}. This happens when two conditions on the network topology and on the fluid conditions at the inlets and outlets of the network meet: 
\begin{enumerate}
\item if there are $N$ inlets and outlets in the network, then fluid flow (Dirichlet) is imposed at least on $N-1$ of them. 
\item there is no loop in the network. 
\end{enumerate}
With these two conditions, velocity profiles and amplitudes in the network are disturbed near the particle but recover when going away from it, going eventually back to the state they have in the absence of particle, fully developed again. The distance for the fluid to become fully developed \cite{Durst} should be correlated to the size of the camera frame, typically the camera size should be at least twice the developed distance. With such size for the camera frame, velocity profiles in the absence of particle  become a very good approximation for Dirichlet fluid boundary conditions on the camera boundary. This gives however few information on pressure distributions and consequently on fluid constraints, these information will however be a result of the numerical simulation.

For example, both conditions 1. and 2. are verified in any channel with Dirichlet (velocity) conditions at either or both extremities, or in any tree-like networks with flow conditions at leaflets. Boundary conditions can  be computed either by an a priori numerical simulation, or by a theoretical calculation of the flow velocities in the network without the particle. For example, under Poiseuille's regime, the velocity profile in a straight channel is parabolic.

Actually, if there are more than one pressure conditions (Neumann) at network inlets or outlets, and/or if there is any loop in the fluid domain, then the particle affects the fluid properties globally. The particle is seen by distant fluid regions as an added pressure drop somewhere in the network and flow rates are re-distributed accordingly to the position and amount of that added pressure drop. Both conditions 1. and 2. are however not compulsory when 1D approximations are available (such as linear regime with Stokes flow), because the fluid properties can be determined by the coupling of the camera method equations with 1D equations that links pressure drops and flow rates in the network with the added pressure drop due to the particle. Since this present work is focused on the camera method itself and is the first to do so, we chose for the sake of clarity to avoid for now such aspects that increase greatly the complexity of the method description. 

The shape of the camera frame for our 2D and 3D examples is spherical (sections \ref{2D} and \ref{3D}), however any shape can be used and they can also change (smoothly) with time and/or particle position. In our 2D axi-symmetric example, the shape of the camera frame depends on the particle position, the camera boundaries coincide with the channel wall (section \ref{risso2Daxi}). Similarly, one can make the camera shape deform to contain at any time the wake of the particle or the boundary layer, if any. Moreover, camera frame shape and boundary conditions can be tuned to mimic phenomenon such as periodicity (section \ref{RBCper}) or insulation.

Finally, if the particle has inertia (heavy particle, high Reynolds number, etc.), then contact with walls are possible. This is not an issue when walls are defined with a penalization method, but convergence problems could occur if time steps are not finely tuned. Actually, if time steps become too small and go to zero then the computation time may increase a lot, on the contrary if time steps are too large, then the particle can "jump" into the wall and be partially "trapped" inside and suffer non physical adhesion or deformation.

\section{Conclusion}

In this work, we propose an original numerical method to track a solid or deformable particle in a fluid network whatever the dimension. The camera method is well adapted as soon as the study needs to focus on the particle behaviour, but not only. With this method, the fluid-structure interaction problem is not solved in the whole fluid domain, and the mesh is limited to a domain whose size is of the order of the size of the particle. The camera method makes also the mesh rotate and translate with the particle to avoid most of the remeshing. In this paper, we focus on the camera method and thus we used a simple fluid background: we make the hypothesis that the fluid velocity in the fluid parts far from the particle is not altered by it. However this hypothesis is not compulsory and can be bypassed by coupling the camera method with 1D fluid models. Consequently, the camera method is very flexible and can be used in a large number of situations and we plan to develop it in future works.

\end{document}